\PassOptionsToPackage{unicode}{hyperref}
\PassOptionsToPackage{hyphens}{url}
\PassOptionsToPackage{dvipsnames,svgnames,x11names}{xcolor}
\documentclass[12pt]{article}
\usepackage{amsmath,amssymb}

\usepackage{graphicx,psfrag,epsf,float}
\usepackage{enumerate}
\usepackage{natbib}
\usepackage{etoolbox}    
\usepackage{url} 
\usepackage[english]{babel}
\usepackage{geometry,theorem,xr,enumitem,xcolor}
\usepackage{setspace}
\usepackage{subfig}
\usepackage{makecell}

\usepackage[abs]{overpic}

\newcommand{\cdummy}{\cdot}

\newcommand{\tmmathbf}[1]{{\boldsymbol{#1}}}
\newcommand{\tmop}[1]{{\operatorname{#1}}}

\newtheorem{proposition}{Proposition}
{\theorembodyfont{\rmfamily}\newtheorem{remark}{Remark}}
\newtheorem{theorem}{Theorem}
\newtheorem{assumption}{Assumption}

\makeatletter
\newcommand*{\addFileDependency}[1]{
  \typeout{(#1)}
  \@addtofilelist{#1}
  \IfFileExists{#1}{}{\typeout{No file #1.}}
}
\makeatother

\newcommand*{\myexternaldocument}[1]{%
    \externaldocument{#1}%
    \addFileDependency{#1.tex}%
    \addFileDependency{#1.aux}%
}

\myexternaldocument{supp} 

\usepackage{iftex}
\ifPDFTeX
  \usepackage[T1]{fontenc}
  \usepackage[utf8]{inputenc}
  \usepackage{textcomp} 
\else 
  \usepackage{unicode-math}
  \defaultfontfeatures{Scale=MatchLowercase}
  \defaultfontfeatures[\rmfamily]{Ligatures=TeX,Scale=1}
\fi
\usepackage{lmodern}
\ifPDFTeX\else  
\fi
\IfFileExists{upquote.sty}{\usepackage{upquote}}{}
\IfFileExists{microtype.sty}{
  \usepackage[]{microtype}
  \UseMicrotypeSet[protrusion]{basicmath} 
}{}
\makeatletter
\@ifundefined{KOMAClassName}{
  \IfFileExists{parskip.sty}{%
    \usepackage{parskip}
  }{
    \setlength{\parindent}{0pt}
    \setlength{\parskip}{6pt plus 2pt minus 1pt}}
}{
  \KOMAoptions{parskip=half}}
\makeatother
\usepackage{xcolor}
\makeatletter
\ifx\paragraph\undefined\else
  \let\oldparagraph\paragraph
  \renewcommand{\paragraph}{
    \@ifstar
      \xxxParagraphStar
      \xxxParagraphNoStar
  }
  \newcommand{\xxxParagraphStar}[1]{\oldparagraph*{#1}\mbox{}}
  \newcommand{\xxxParagraphNoStar}[1]{\oldparagraph{#1}\mbox{}}
\fi
\ifx\subparagraph\undefined\else
  \let\oldsubparagraph\subparagraph
  \renewcommand{\subparagraph}{
    \@ifstar
      \xxxSubParagraphStar
      \xxxSubParagraphNoStar
  }
  \newcommand{\xxxSubParagraphStar}[1]{\oldsubparagraph*{#1}\mbox{}}
  \newcommand{\xxxSubParagraphNoStar}[1]{\oldsubparagraph{#1}\mbox{}}
\fi
\makeatother

\usepackage{longtable,booktabs,array}
\usepackage{calc} 
\usepackage{etoolbox}
\makeatletter
\patchcmd\longtable{\par}{\if@noskipsec\mbox{}\fi\par}{}{}
\makeatother
\IfFileExists{footnotehyper.sty}{\usepackage{footnotehyper}}{\usepackage{footnote}}
\makesavenoteenv{longtable}
\usepackage{graphicx}
\makeatletter
\def\maxwidth{\ifdim\Gin@nat@width>\linewidth\linewidth\else\Gin@nat@width\fi}
\def\maxheight{\ifdim\Gin@nat@height>\textheight\textheight\else\Gin@nat@height\fi}
\makeatother
\setkeys{Gin}{width=\maxwidth,height=\maxheight,keepaspectratio}
\makeatletter
\def\fps@figure{htbp}
\makeatother

\makeatletter
\@ifpackageloaded{caption}{}{\usepackage{caption}}
\AtBeginDocument{%
\ifdefined\contentsname
  \renewcommand*\contentsname{Table of contents}
\else
  \newcommand\contentsname{Table of contents}
\fi
\ifdefined\listfigurename
  \renewcommand*\listfigurename{List of Figures}
\else
  \newcommand\listfigurename{List of Figures}
\fi
\ifdefined\listtablename
  \renewcommand*\listtablename{List of Tables}
\else
  \newcommand\listtablename{List of Tables}
\fi
\ifdefined\figurename
  \renewcommand*\figurename{Figure}
\else
  \newcommand\figurename{Figure}
\fi
\ifdefined\tablename
  \renewcommand*\tablename{Table}
\else
  \newcommand\tablename{Table}
\fi
}
\@ifpackageloaded{float}{}{\usepackage{float}}
\floatstyle{ruled}
\@ifundefined{c@chapter}{\newfloat{codelisting}{h}{lop}}{\newfloat{codelisting}{h}{lop}[chapter]}
\floatname{codelisting}{Listing}

\makeatother
\makeatletter
\@ifpackageloaded{caption}{}{\usepackage{caption}}
\@ifpackageloaded{subcaption}{}{\usepackage{subcaption}}
\makeatother

\ifLuaTeX
  \usepackage{selnolig}  
\fi
\usepackage[]{natbib}
\usepackage{bookmark}

\IfFileExists{xurl.sty}{\usepackage{xurl}}{} 
\hypersetup{
  pdftitle={Title},
  pdfauthor={Author 1; Author 2},
  pdfkeywords={3 to 6 keywords, that do not appear in the title},
  colorlinks=true,
  linkcolor={blue},
  filecolor={Maroon},
  citecolor={Blue},
  urlcolor={Blue},
  pdfcreator={LaTeX via pandoc}}

\newcommand{\anon}{1}

\begin{document}

\def\spacingset#1{\renewcommand{\baselinestretch}%
{#1}\small\normalsize} \spacingset{1}


\if1\anon
{
  \title{\bf Consistent Bayesian Local Spatial Feature Selection with Application to Spatial Multimodal Omics}
  \author{Kun Huang$^{1}$, Xiyu Peng$^{1}$, Huiyan Sang$^{1}$, and Ligang Lu$^{2}$
   \hspace{1cm}\\
   \small{
    $^{1}$Department of Statistics, Texas A\&M University, College Station, TX, USA}\\
    \small{$^{2}$Shell International Exploration and Production Inc, Houston, TX, USA}    }
  \date{}
  \maketitle
} \fi

\if0\anon
{
  \bigskip
  \bigskip
  \bigskip
  \begin{center}
    {\LARGE\bf Consistent Bayesian Local Spatial Feature Selection with Application to Spatial Multimodal Omics}
\end{center}
  \medskip
} \fi

\bigskip
\begin{abstract}
Motivated by a high-dimensional regression problem in spatial multimodal omics (SMO), we propose a Bayesian framework for local spatial feature selection, where a random domain partition prior is introduced to divide the spatial domain into several contiguous clusters with flexible shapes and an unknown number of clusters, conditional on which a local feature selection prior is imposed within each cluster. The notion of “feature” is general and may include both covariates and functional bases, allowing the framework to perform both local variable selection and local basis selection, the latter being essential for adaptively approximating spatially varying functions with localized characteristics. We derive coupled hyperparameter conditions linking domain partition and local feature selection priors, under which the consistency theory and posterior contraction rates of both the domain partition and feature selection are established. 
We develop an efficient informed reversible jump Markov chain Monte Carlo algorithm to address the computational challenges encountered in joint posterior sampling of domain partitions and selected features.  Simulation studies demonstrate the effectiveness of the proposed model and algorithm, highlighting its advantages over existing methods. The application of our model to an SMO dataset reveals biologically meaningful spatial patterns within breast cancer tissue.
\end{abstract}

\noindent%
\vfill

\newpage
\spacingset{1.8} 

\section{Introduction}\label{SEC:intro}
\subsection{Motivation}
Spatial transcriptomics (ST) is a recently developed and rapidly advancing technology that enables simultaneous measurement of gene expression and spatial localization across tissue sections \citep{yan2024bayesian}. On the grounds of ST, spatial multimodal omics (SMO), an approach that enables simultaneous detection of gene expression and other types of omics modalities  (e.g., proteins, metabolite profiles) within the same tissue spatial context, is crying out to emerge and becoming the next hot topic \citep{yue2023guidebook}. Our research is motivated by a publicly available SMO dataset published by \citet{10xgenomics_breastcancer}, which includes 4169 spatial spots from a breast cancer tissue sample (see Figure~\ref{FIG:STcluster}(a) for the tissue image). Each spot contains expression measurements for the whole transcriptome, covering 17957 genes, along with 35 selected protein markers. 

While existing methods aim to integrate multiple modalities to delineate distinct spatial domains~\citep{Coleman2025}, our interests lie in quantifying the relationship between different modalities—for instance, between proteins and RNAs within a spatial context.
The interaction between protein and RNA in tumor microenvironments has recently attracted growing attention (e.g., \citealp{arad2023functional,ban2024protein}). 
In this study, we focus on the CD8A protein, which is a defined protein marker of CD8$^{+}$ T cells, a key player of anti-tumor immunity, and study its relationship with gene expression.

This problem poses three main challenges. 
First, although 17957 genes are measured, only a small subset is functionally relevant to CD8$^{+}$ T cells, making variable selection a critical component of the analysis.
Second, due to the complexity of the tumor microenvironment, protein–gene relationships can vary spatially across different tissue domains (e.g., tumor regions, stromal regions). In particular, in cancer samples, omics features often exhibit abrupt changes at tumor boundaries \citep{chen2024investigating}, which may further lead to a sharp change in protein–gene relationships. 
Third, the large number of spatial spots and measured genes imposes a substantial computational burden. 
To overcome these challenges, below we propose a scalable spatially varying coefficient regression 
model for large-scale data that simultaneously enables gene selection and captures spatial heterogeneity, including sharp changes, across different tissue domains. 

Specifically, let $\{\tmmathbf{s}_i, \tmmathbf{x}(\tmmathbf{s}_i),y (\tmmathbf{s}_i)\}_{i = 1}^n$ be the data at $n$ observed spatial locations $\tmmathbf{s}_1, \ldots, \tmmathbf{s}_n \in \mathcal{D}
\subset \mathbb{R}^2$, where $\tmmathbf{x}(\tmmathbf{s}_i)$ and $y (\tmmathbf{s}_i)$ are the feature and response variable at $\tmmathbf{s}_i$, respectively. In our SMO dataset, $\tmmathbf{x}(\tmmathbf{s}_i)$ and $y (\tmmathbf{s}_i)$ represent the gene expression measurements and the CD8A protein expression at $\tmmathbf{s}_i$, respectively. We consider a regression model $y (\tmmathbf{s}_i) = \mu (\tmmathbf{s}_i) + \epsilon (\tmmathbf{s}_i)$, where $\mu (\cdot)$ is the mean function of $y(\cdot)$, and $\{ \epsilon (\tmmathbf{s}_i)
\}_{i = 1}^n$ are independent Gaussian noises with mean 0 and $\sup_{1
\leqslant i \leqslant n} \mathbb{E} \epsilon^2 (\tmmathbf{s}_i) < \infty$. In this paper, we study the relationship between $\mu(\cdot)$ and $\tmmathbf{x}(\cdot)$ by a spatially varying coefficient model 
\begin{equation*}
\mu (\tmmathbf{s}_i)=\tmmathbf{x}^{T}(\tmmathbf{s}_i) \tilde{\tmmathbf{\theta}}(\tmmathbf{s}_i),
\end{equation*}
where $\tilde{\tmmathbf{\theta}}(\cdot)$ is a spatially varying regression coefficient, and we use the notation $\tilde{\cdot}$ to indicate that $\tilde{\tmmathbf{\theta}}(\cdot)$ is sparse with finitely many nonzero entries. 
To account for spatial heterogeneity and capture potential sharp changes across tissue subregions, we assume $\tilde{\tmmathbf{\theta}}(\cdot)$ to be a piecewise constant function on domain $\mathcal{D}$, i.e., the domain $\mathcal{D}$ can be partitioned into some clusters, such that $\tilde{\tmmathbf{\theta}} (\tmmathbf{s}_i) =\tilde{\tmmathbf{\theta} }(\tmmathbf{s}_{i'})$ if
$\tmmathbf{s}_i$ and $\tmmathbf{s}_{i'}$ belong to the same cluster, and $\tilde{\tmmathbf{\theta}} (\tmmathbf{s}_i) \neq \tilde{\tmmathbf{\theta}}
(\tmmathbf{s}_{i'})$ otherwise. 
The sparsity pattern of $\tilde{\boldsymbol{\theta}}(\cdot)$ may vary across subregions, as different subsets of features may be associated with the response in different subregions. We refer to this model setting as a local spatial feature selection (LSFS) model throughout the paper. 

Under the LSFS framework, we focus on two key objectives. First, to identify “local subregions” (clusters) through a domain partitioning approach; second, to perform local feature selection of nonzero entries in $\tilde{\tmmathbf{\theta}}(\cdot)$ based on the resulting partition. LSFS naturally supports spatial predictions at unobserved locations. It also accommodates a broader class of model settings and applications, as features need not be limited to observed covariates; they can also be constructed or derived, for example, using functional basis expansions as in nonparametric or semiparametric regression models. In such settings, the LSFS framework facilitates local selection of basis functions, allowing the model to adaptively approximate latent spatially varying functions that exhibit local characteristics, such as varying smoothness.  





\subsection{Related work}\label{SEC: Related works}

\noindent\textbf{Domain partition methods.} Most existing spatial clustering methods consider clustering the $n$ observed locations into some disjoint sub-clusters~\citep{page_spatial_2016, quintana2022dependent,hu2023bayesian}. These approaches are not suitable for our SMO dataset, as our primary interest lies in partitioning the tissue region rather than only the observed $n$ locations. Alternatively, a model-based domain partition model considers a random partition of the entire domain $\mathcal{D}$ into some disjoint sub-domains, which is more appropriate for our analysis. Popular domain partition models include binary decision trees \citep{denison1998bayesian}, Voronoi tessellation models \citep{knorr2000bayesian}, and boundary curve models~\citep{masotti2024general}. These models impose some restrictive constraints on the cluster boundary shapes and struggle to handle domains with physical barriers. Existing theoretical work of spatially clustered regression model mostly focuses on 
showing that the posterior is a consistent estimate of the true regression parameter or data-generating density ~\citep{luo2021bayesian}, while the more relevant problem of clustering consistency remains largely unexplored due to its technical challenges. 
Very recently, \cite{shenweining,zheng2024consistency,huangandsang2025consistency} developed Bayesian clustering consistency theory for spatial and graph-structured dependent data. However, none of these consider local variable selection.

\noindent\textbf{Spatial variable selection methods.} A number of methods have been proposed for spatial variable selection problems \citep[see, e.g., ][]{zhu2010selection,xie2019spatial}, where the regression coefficients of selected spatial variables are assumed to be constant across space. \cite{li2021sparse} and \citet{dambon2022joint} selected spatial variables while allowing spatially varying effects, based on spline models and Gaussian processes, respectively.
These methods assume global variable selection in the sense that the inclusion or exclusion of a covariate is consistent across the entire spatial domain.  
Recently, the idea of ``local variable selection" 
has been receiving increasing attention. \citet{dai2025moving} highlighted the importance of local variables. 
\cite{rossell2025semiparametric} studied local variable selection for semi-parametric models, where variables are selected at some given local regions, and establishes variable selection consistency. 
 \cite{boehm2015spatial} proposed a Bayesian spatial variable selection method, allowing both local and global variable selection, but its computation can only handle a few hundred spatial locations. None of the above approaches performs local variable selection by identifying local subregions and selecting variables simultaneously.  \cite{zhong2023sparse} proposed a frequentist framework to cluster the observed locations and select variables simultaneously. However, their method does not partition the domain, lacks uncertainty quantification, and does not guarantee clustering consistency.

\noindent\textbf{Adaptive function approximation.} In nonparametric regression, adaptive function approximation has a long history. \cite{fan1995data} proposed variable bandwidth selection for kernel regression. \cite{zhou2001spatially} and \cite{yuan2013adaptive} investigated adaptive knot placement in spline models. Functional basis selection also plays an important role in adaptive approximation. For example, \cite{sklar2013nonparametric} employed cross-validation to select bases from predefined libraries, and \cite{meng2022smoothing} studied basis selection in smoothing spline approximation. However, none of the existing approaches performs local basis selection by identifying local subregions and selecting bases simultaneously. This problem is particularly complex in spatial contexts, where the identification of local subregions is challenging, especially in domains with complex geometries.

\subsection{Contributions}
We propose a Bayesian framework to model the spatially varying relationship between $\tmmathbf{x}(\cdot)$ and $y(\cdot)$ with LSFS. For domain partitioning, we discretize $\mathcal{D}$ into blocks and define a random domain partition prior through a spanning tree-based graph partition construction. This method flexibly models contiguous, irregularly shaped clusters while respecting domain geometries, and the blocking reduces the infinite domain partition space to
the finite blocking partition space, lowering the computational burden and making partitioning feasible and scalable. Conditional on the domain partition, we propose a prior to select features locally by penalizing the number of selected features. 


Both domain partitioning and feature selection are challenging high-dimensional tasks, and LSFS, rather than being a straightforward combination of the two, presents two major challenges beyond those encountered when addressing each task individually. The first challenge lies in prior specification, particularly the hyperparameter selection strategy, which plays a central role in high-dimensional problems. Notably, the domain partitioning and feature selection are interdependent. On one hand, feature selection is performed conditional on the domain partition and thus depends on it. On the other hand, the feature selection prior also affects the resulting domain partition. For example, if the feature number penalty is too strong, such that some true local features are excluded, the model may incorrectly favor a partition with more clusters in an attempt to better fit the data. Consequently, the priors for domain partitioning and feature selection must be carefully jointly designed to ensure consistent estimation of both components. 
We conduct theoretical analysis and establish consistency results for simultaneous domain partitioning and feature selection, which also provide theoretical guidance for joint prior hyperparameter selection. To the best of our knowledge, this is the first work considering both tasks simultaneously with guaranteed consistency. 


The second challenge lies in computation. In SMO datasets, the numbers of spatial locations $n$ and features can both be large, resulting in a high-dimensional joint space of domain partitions and selected features. This complexity leads to severe slow mixing issues in standard reversible jump Markov chain Monte Carlo (RJ-MCMC) methods commonly used for this type of trans-dimensional model selection problem.
To address this issue, we propose an informed RJ-MCMC algorithm that enables more efficient joint updates of domain partitions and selected features. Simulation studies demonstrate that the proposed informed RJ-MCMC mixes well and is computationally efficient.

The rest of the paper is organized as follows. Sections \ref{SEC:model} and \ref{SEC:computation} introduce our model and the proposed informed RJ-MCMC algorithm, respectively. Section \ref{SEC:theory} presents theoretical results. In Section \ref{SEC:simu}, we conduct numerical simulations to demonstrate the effectiveness of the proposed model and algorithm, as well as the comparison with some existing methods. In Section \ref{SEC:realdata}, we apply our model to the SMO dataset and demonstrate results. Technical proofs and computational details are contained in Supplementary Material.

\section{Model}\label{SEC:model}
As mentioned in the Introduction, we consider the model 
  $y (\tmmathbf{s}_i) = \tmmathbf{x}^{T}(\tmmathbf{s}_i) \tilde{\tmmathbf{\theta}}(\tmmathbf{s}_i)+ \epsilon (\tmmathbf{s}_i)$,
where the feature $\tmmathbf{x}(\cdot)$ can include both covariates and functional bases. Generally, we write
$\tmmathbf{x}^T (\cdot) = \{ \tmmathbf{x}^T_0 (\cdot),
\tmmathbf{x}^T_{ \dagger} (\cdot) \}$, where $\tmmathbf{x}_0 (\cdot)$ denotes features believed to be predictive and included without selection, and $\tmmathbf{x}_{ \dagger} (\cdot)$ denotes features subject to selection. 
We now elaborate on three spatial model settings under the framework of LSFS. 
\begin{itemize}
\item \textbf{Variable selection (VS)}. In this setting, $\tmmathbf{x}_{ \dagger} (\cdot)$ is a high-dimensional covariate and we set $\tmmathbf{x}_0 (\cdot) \equiv 1$ as an intercept term.  
We aim to select the subset of important covariates from $\tmmathbf{x}_{ \dagger} (\cdot)$ that is predictive of $\mu(\cdot)$ locally. 
\item \textbf{Semi-parametric variable selection (SVS)}. This setting is similar to VS except that
we include 
functional bases of $\tmmathbf{s}$ (e.g., splines or polynomial bases) in
$\tmmathbf{x}_0 (\cdot)$ to model the spatially smoothly varying random effects that cannot be explained by the linear model of
$\tmmathbf{x}_{ \dagger} (\cdot)$. 

\item \textbf{Nonparametric regression (NR)}. In this setting, we set $\tmmathbf{x}_0 (\cdot) \equiv 1$, and include functional bases of $\tmmathbf{s}$ as features in $\tmmathbf{x}_{
\dagger} (\cdot)$. For example, if we consider polynomial bases of degree two, we write $\tmmathbf{x}^T_{
\dagger} (\tmmathbf{s})=(s_1,s_2,s_1^2,s_1s_2,s_2^2)$, where $\tmmathbf{s} = (s_1, s_2)$.
We aim to select ``appropriate" functional bases from $\tmmathbf{x}_{
\dagger} (\cdot)$ in a locally adaptive way to learn a nonparametric spatial latent function $\mu(\tmmathbf{s})$ for prediction at new locations. 
The rationale is from the Taylor expansion: for functions with higher-order derivatives, higher-degree polynomials are preferred for approximation, as they yield smaller approximation bias. In particular, 
$\mu(\cdot)$ may be approximately linear in some regions, exhibit quadratic behavior in others, or vary in smoothness across locations, thereby necessitating the use of location-specific bases. 
\end{itemize}

 


We propose a unified Bayesian framework for joint domain partitioning and feature selection under the three settings described above. Let $\pi(\mathcal{D})$ denote a partition of domain $\mathcal{D}$, and $\pi (j)$ be
the set of observed locations in its $j$-th cluster, where $1\leq j \leq k$ and $k$ is the number of clusters. Let $q$ be the dimension of $\tmmathbf{x}_{\dagger} (\cdot)$ and write $\nu \in \mathcal{I}= \{
0, 1 \}^q$ as a $q$-dimensional vector with entries being either $1$ or $0$, indicating if the feature at the corresponding position is selected or not. For convenience, we adopt a dualistic view of $\nu$ as both a binary
indicator vector, and a subset of $\{ 1, \ldots, q \}$ indicating nonzero entries. We also refer to $\nu$ as the ``active set'' in the following context. 
Write
$\nu_j \in \mathcal{I}$ as the active set for the $j$-th cluster and
$\tmmathbf{\nu} = \{ \nu_j \}_{j = 1}^k$.
Conditional on $\pi(\mathcal{D})$ and $\tmmathbf{\nu}$, we assume $\tilde{\tmmathbf{\theta}}(\cdot)$ to be a constant within each cluster. For the $j$-th cluster, let $\tmmathbf{\theta}_j$ denote the subvector of $\tilde{\tmmathbf{\theta}}(\cdot)$ corresponding to the selected feature $\tmmathbf{x}_{\nu_j}(\cdot)$, and $\tmmathbf{\theta} =
 \{ \tmmathbf{\theta}_j \}_{j = 1}^k$, where $\tmmathbf{x}^{T}_{\nu}(\cdot)=\{\tmmathbf{x}^{T}_{0}(\cdot),\tmmathbf{x}^{T}_{\nu,\dagger}(\cdot)\}$, and $\tmmathbf{x}_{\nu,\dagger} (\cdot)$ is the selected subvector of $\tmmathbf{x}_{\dagger} (\cdot)$ for a given $\nu$. Write $\mathcal{X}$ as the $\sigma$-algebra generated by $\tmmathbf{x}(\cdot)$ and $\mathcal{S} = \{\tmmathbf{s}_i\}_{i=1}^n$. For a location set $\mathcal{S}_0
\subseteq \mathcal{S}$, we denote $\tmmathbf{x}_{\mathcal{S}_0, \nu}$ as the corresponding design matrix generated by $\{ \tmmathbf{x}_{\nu}
(\tmmathbf{s}), \tmmathbf{s} \in \mathcal{S}_0 \}$, and $\tmmathbf{y}_{\mathcal{S}_0} = \{ y 
(\tmmathbf{s}), \tmmathbf{s} \in \mathcal{S}_0 \}$. Write $\tmmathbf{y} = \{{y}(\tmmathbf{s}_i)\}_{i=1}^{n}$.
The data likelihood model is 
\begin{equation}
  \mathbb{P} \{\tmmathbf{y} \mid \pi(\mathcal{D}), \tmmathbf{\nu},
  \tmmathbf{\theta}, \mathcal{S}, \mathcal{X}\} = \prod_{j = 1}^k
  \mathbb{P}_{\tmop{Gaussian}} \{ \tmmathbf{y}_{\pi (j)} ;
  \tmmathbf{x}_{\pi (j), \nu_j} \tmmathbf{\theta}_j, \sigma^2
  \tmmathbf{I}_{| \pi (j) |} \}, \label{EQ:lly}
\end{equation}
where $\mathbb{P}_{\tmop{Gaussian}} (\cdot ; \tmmathbf{a},
\tmmathbf{\Sigma})$ denotes the probability density function of the Gaussian
distribution with mean $\tmmathbf{a}$ and covariance $\tmmathbf{{\Sigma}}$, 
$\sigma^2$ is a pre-specified constant for the variance of noise, $\tmmathbf{I}_a$
is an $a \times a$ identity matrix for a given integer $a$, and $| \cdot |$ denotes the cardinality.
We fix $\sigma^2$ at a constant value in our model, which reduces the model space and facilitates computation. In our theoretical analysis (Section \ref{SEC:theory}), we show that the consistency results remain valid for any fixed constant value of $\sigma^2$. Therefore, the pre-specified $\sigma^2$ is not required to equal the true variance of $\epsilon (\cdot)$, which itself is not assumed to be constant across locations.  

\subsection{A prior model for domain partition $\pi(\mathcal{D})$}\label{subsec:spanningtree}

 Following the similar domain partition model in \cite{huangandsang2025consistency}, we model $\pi(\mathcal{D})$ upon a random graph partition prior on a discretized mesh of $\mathcal{D}$. To simplify theoretical analysis, we assume $\mathcal{D}= [0, 1]^2$ and segment $\mathcal{D}$ into $K^2$ disjoint regular gridded blocks, $\{B_m \}_{m = 1}^{K^2}$. 
In Section~\ref{SEC:theory}, we provide
a rate condition of $K$ to guarantee domain partition consistency. 
The method can adopt other mesh blocking strategies in practice, and theoretical results can be extended to a more general domain homeomorphic to $[0, 1]^2$  under the Euclidean metric with a bi-Lipschitz mapping.

\begin{figure}[h]
		\centering
		\begin{tabular}{ccc}
			{\includegraphics[width=0.3\linewidth,height=0.2\textheight]{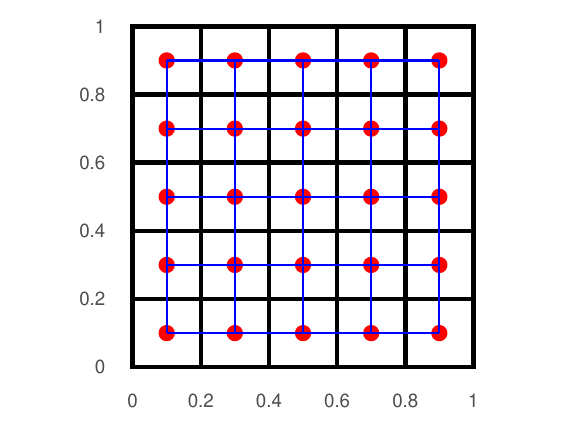}}&
			{\includegraphics[width=0.3\linewidth,height=0.2\textheight]{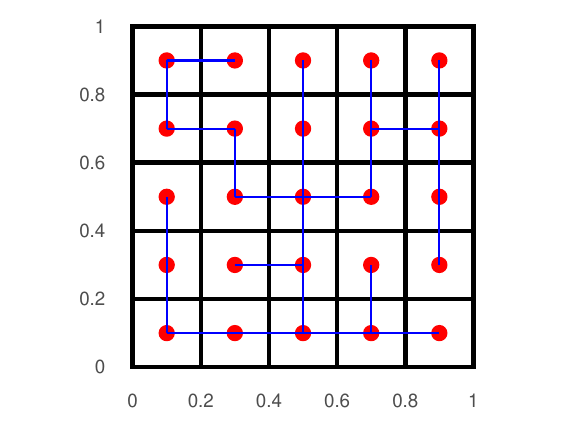}}&
			{\includegraphics[width=0.3\linewidth,height=0.2\textheight]{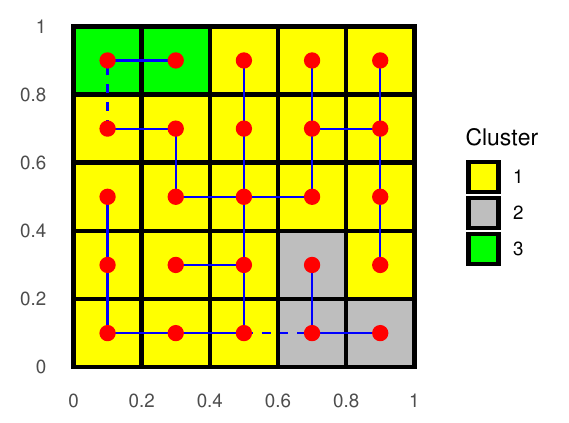}}
			\\
			{\small (a)} & {\small (b)} &{\small (c) } 
		\end{tabular}
        
		\caption{Illustration of our partition model with $K=3$. (a) Graph $\mathcal{G}=
(\mathcal{V}, \mathcal{E})$, where edges in $\mathcal{E}$ are denoted by the blue lines between adjacent blocks. (b) One spanning tree of the graph $\mathcal{G}$ in (a). (c) Domain partition $\pi^\ast$ induced by cutting two edges (dashed lines) of the spanning tree in (b). Each cluster (denoted by different colors) is a connected component. Locations within the same block have the same cluster membership.}
		\label{FIG:illustration}
        \vspace{-10pt}
	\end{figure}

We construct an undirected spatial graph, $\mathcal{G}= (\mathcal{V}, \mathcal{E})$, where $\mathcal{V}= \{B_m \}_{m = 1}^{K^2}$ is the set of vertices, and $\mathcal{E}$ is the set of
edges connecting only adjacent blocks (see Figure \ref{FIG:illustration}(a)
for the constructed $\mathcal{G}$).
Clustering the vertices naturally induces a partition of $\mathcal{D}$. In spatial applications, it is often desirable to enforce spatial contiguity in this partition, so that the resulting clusters can be more naturally interpreted as spatial subregions. 

We formally define contiguous partitions as follows. 
Given an undirected graph $\mathcal{G}$, a subset
$\mathcal{V}_0 \subseteq \mathcal{V}$ is a contiguous cluster if there exists
a connected subgraph $\mathcal{G}_0 = (\mathcal{V}_0, \mathcal{E}_0)$, where
$\mathcal{E}_0 \subseteq \mathcal{E}$. We say $\pi (\mathcal{V}) =
\{\mathcal{V}_1, \ldots, \mathcal{V}_k \}$ is a contiguous partition of
$\mathcal{V}$ with respect to $\mathcal{G}$, if $\mathcal{V}_j \subseteq
\mathcal{V}$ is a contiguous cluster for $j = 1, \ldots, k$, $\cup_{j = 1}^k
\mathcal{V}_j =\mathcal{V}$, and $\mathcal{V}_j \cap \mathcal{V}_{j'} = \emptyset$
for $j \neq j'$. 

Let $\mathcal{P}_G$ denote the space of all possible contiguous partitions with respect to the graph $\mathcal{G}$. A Bayesian spatial domain partition model amounts to specifying a prior distribution supported on the discrete space $\mathcal{P}_G$, which is usually a challenging task due to its large and complex combinatorial structure. 
Recently, several model-based spanning tree partitioning models have been proposed to model spatially contiguous clusters. A spanning tree of $\mathcal{G}$ is a subgraph $\mathcal{T}=
(\mathcal{V}, \mathcal{E}_{\mathcal{T}})$, where 
$\mathcal{E}_{\mathcal{T}} \subseteq \mathcal{E}$ connects all vertices without any cycles. See Figure \ref{FIG:illustration}(b) for an example of a spanning tree $\mathcal{T}$ induced from $\mathcal{G}$. Removing $k-1$ edges from a spanning tree induces a contiguous partition with $k$ clusters with respect to $\mathcal{G}$. This nice property motivates a hierarchical construction of a random domain partition prior through random spanning trees. Specifically, the model proceeds by first assigning a prior model, $\mathbb{P}(\mathcal{T})$, for $\mathcal{T}\in \Delta$, where $\Delta$ represents the space of all possible spanning trees of $\mathcal{G}$. Conditional on $\mathcal{T}$, we then place a random partition prior model, $\mathbb{P}\{\pi(\mathcal{V})|\mathcal{T}\}$, for $\pi(\mathcal{V}) \in \mathcal{P}_{\mathcal{T}}$, where $\mathcal{P}_{\mathcal{T}}$ represents the space of contiguous partitions with respect to $\mathcal{T}$.

\noindent
\textbf{Choices of} $\mathbb{P}(\mathcal{T})$:
Uniform spanning tree prior \citep[UST,][]{teixeira2019bayesian} is a popular way to model spanning trees, which assumes that $\mathbb{P}(\mathcal{T}) \propto 1$.  
Another popular spanning tree model is the random minimum spanning tree prior \citep[RST,][]{luo2021bayesian}. Specifically, for any given weighted graph, 
a minimum spanning tree (MST) is a spanning tree whose total edge weights is the smallest among all possible spanning trees of the graph. 
RST is a generative prior defined as 
\begin{equation}\label{eq:tree_prior}
\begin{aligned}
& w_e  \stackrel{iid}\sim \tmop{Unif}(0, 1), \text{where $w_e$ is the weight assigned to the $e$-th edge of $\mathcal{G}$} \\
&\mathcal{T}_{\boldsymbol{w}} = \text{MST}(\boldsymbol{w}), \text{where $\boldsymbol{w} = \left\{w_e \right\}_{e\in \mathcal{E}}$} 
\end{aligned}
\end{equation}
Any spanning tree of $\mathcal{G}$ can be the result of an MST for certain $\boldsymbol{w}$~\citep{luo2021bayesian}.  

\noindent
\textbf{Choices of} $\mathbb{P}\{\pi(\mathcal{V})|\mathcal{T}\}$:
We assume the number of clusters $k$ follows a truncated
Poisson distribution with mean parameter $\lambda$:
\begin{equation}
  k \sim \tmop{Poisson} (\lambda) \cdot \mathbb{I} (1 \leqslant k \leqslant
  k_{\max}), \label{DEF:k}
\end{equation}
where $k_{\max}$ is a pre-specified maximum number of clusters, and $\mathbb{I}(\cdot)$ is the indicator function. Conditional on
$\mathcal{T}$ and $k$, we assume:
\begin{equation}
  \mathbb{P} \{\pi (\mathcal{V}) |k, \mathcal{T}\} \propto \mathbb{I} \left\{
  \pi (\mathcal{V}) \text{ is induced from } \mathcal{T} \text{ and has } k
  \text{ clusters} \right\} . \label{DEF:pi}
\end{equation}

$\pi(\mathcal{V})$ immediately leads to a domain partition by assigning all locations within the same block to the same cluster (see Figure \ref{FIG:illustration}(c)). The clustering of the observed locations, $\mathcal{S}$, is obtained accordingly. In what follows, we write $\pi^{\ast}$ as the domain partition induced by $\pi(\mathcal{V})$, $\pi^\ast(j)$ as the set of observed locations in the $j$-th clsuter of $\pi^\ast$, and $\pi$ as the corresponding clustering of $\mathcal{S}$.

\subsection{Priors for feature selection and regression coefficients}\label{SEC:featureselection}


Conditional on $(\pi^\ast,k,\mathcal{T},\mathcal{S},\mathcal{X})$, we assume the prior of $\tmmathbf{\nu} = \{ \nu_j \}_{j = 1}^k$ as
\begin{equation}
  \mathbb{P} (\tmmathbf{\nu} \mid \pi^{\ast}, k, \mathcal{T}, \mathcal{S},
  \mathcal{X}) = \prod_{j = 1}^k \mathbb{P} \{ \nu_j \mid \pi^{\ast} (j) \},
  \label{EQ:priornu}
\end{equation}
where $\mathbb{P} \{ \nu_j \mid \pi^{\ast} (j) \}  =\mathbb{P} \{ \nu_j \mid |\pi^{\ast} (j)| \} \propto \exp [- | \nu_j |
\alpha \{ | \pi^{\ast} (j) | \}] \mathbb{I} (| \nu_j | \leqslant q_{\max})$
for some positive function $\alpha (\cdot)$ and a pre-specified constant
$q_{\max}$ to upper bound the number of active features, $| \nu_j |$. The function $\alpha (\cdot)$ controls the strength of the penalty for the number of active features 
to 
avoid overfitting. 
In the
classic variable selection literature where no clustering is
considered, the penalty typically increases with sample size (e.g., \citealp{zhao2006model}). Accordingly, we incorporate local sample size information (i.e., $| \pi^{\ast} (j) |$) into $\alpha(\cdot)$ to ensure local adaptivity. In Section \ref{SEC:theory}, we provide a condition of $\alpha (\cdot)$, such that predictive features can be selected consistently under the VS setting. 


Recall that, conditional on $\pi^\ast$ and $\tmmathbf{\nu}$, we assume $\tilde{\tmmathbf{\theta}}(\cdot)$ to be a constant within each cluster. Let $\cdot^\dagger$ denote the pseudoinverse.
 To enable localized shrinkage in variable selection, we adopt a local version of Zellner’s  g-prior, defined as follows:
\begin{equation}
  \mathbb{P} (\tmmathbf{\theta} \mid \pi^{\ast}, k, \mathcal{T},
  \tmmathbf{\nu}, \mathcal{S}, \mathcal{X}) = \prod_{j = 1}^k
  \mathbb{P}_{\tmop{Gaussian}} [\tmmathbf{\theta}_j ; \tmmathbf{0}, \gamma n \sigma^2
  \{ \tmmathbf{x}_{\pi^{\ast} (j), \nu_j}^T \tmmathbf{x}_{\pi^{\ast} (j),
  \nu_j} \}^{\dagger}], \label{EQ:priortheta}
\end{equation}
where $\gamma > 0$ is a pre-specified constant controlling the prior variance.


\section{Informed RJ-MCMC algorithm for Bayesian inference}\label{SEC:computation}

We use MCMC to sample from the posterior distribution $\mathbb{P} (\pi^{\ast},
k, \mathcal{T}, \tmmathbf{\nu}, \tmmathbf{\theta} \mid \mathcal{X},
\mathcal{S}, \tmmathbf{y})$. 
This can be formulated as a model selection inference problem where 
$(\pi^{\ast},
k, \mathcal{T})$ determines the domain partition model, and $(\tmmathbf{\nu}, \tmmathbf{\theta})$ are within-model parameters whose number of dimensions depends on the partition model. 
Thanks to the use of the conjugate Gaussian prior for $\tmmathbf{\theta}$, we can easily derive the Gaussian posterior conditional distribution, $\mathbb{P} (\tmmathbf{\theta} \mid \pi^{\ast}, k, \mathcal{T},
   \tmmathbf{\nu}, \mathcal{X}, \mathcal{S}, \tmmathbf{y})$, and the closed-form collapsed data marginal likelihood $\mathbb{P}(\tmmathbf{y} \mid \pi^{\ast}, k,\mathcal{T},\tmmathbf{\nu}, \mathcal{X}, \mathcal{S})$ integrating out $\tmmathbf{\theta}$. Details are deferred to Supplementary Material.  
However, unlike \(\tmmathbf{\theta}\), the parameters \(\tmmathbf{\nu}\) cannot be easily integrated out, and trans-dimensional sampling methods such as RJ-MCMC are typically required for the joint proposal and update of $(\pi^{\ast},k, \mathcal{T},\tmmathbf{\nu}\mid \mathcal{X}, \mathcal{S}, \tmmathbf{y})$. 

Unfortunately, standard RJ-MCMC poses significant challenges for our model. The number of possible partition models increases exponentially with the number of blocks, which diverges as $n \rightarrow \infty$ according to our theoretical analysis in Section~\ref{SEC:theory}. Standard random birth-death type of samplers often encounter slow mixing issues in this context.  Moreover, the within-model parameter $\nu_j$ in each subregion involves another high-dimensional and trans-dimensional variable selection problem. Applying standard birth-death methods simultaneously to propose
$\pi^{\ast}$ and $\boldsymbol{\nu}$ from their substantially larger joint parameter space would further aggravate the slow mixing problem.

To address this issue, we propose an informed RJ-MCMC approach to sample  
$\pi^{\ast}, k, \mathcal{T}, \tmmathbf{\nu} \mid \mathcal{X},
\mathcal{S}, \tmmathbf{y}$. 
Specifically, let $\pi^\ast=\{\mathcal{D}_1^\ast,\ldots,\mathcal{D}_k^\ast\}$ be the current domain partition with $k$ clusters. We follow one of four moves to propose a new state, 
$(\pi^{\ast}_{\tmop{new}}, k_{\tmop{new}}, \mathcal{T}_{\tmop{new}},\tmmathbf{\nu}_{\tmop{new}})$, 
from the current
state, $(\pi^{\ast}, k, \mathcal{T},\tmmathbf{\nu})$: birth, death, change, and hyper, with
probabilities $r_b (k), r_d (k), r_c (k) \text{ and } r_h (k)$, respectively. 
For the birth move, we randomly
split one cluster in the current $\pi^{\ast}$, say $\mathcal{D}^{\ast}_j$,
into two clusters, say $\mathcal{D}_{\tmop{new}, j 1}^{\ast}$ and
$\mathcal{D}_{\tmop{new}, j 2}^{\ast}$, by randomly removing an edge
connecting two blocks in $\mathcal{D}^{\ast}_j$. Write new active sets of $\mathcal{D}_{\tmop{new}, j 1}^{\ast}$ and
$\mathcal{D}_{\tmop{new}, j 2}^{\ast}$ as $\nu_{\tmop{new}, j 1}$ and
$\nu_{\tmop{new}, j 2}$, respectively. 
We first randomly select one
of \ $(\nu_{\tmop{new}, j 1}, \nu_{\tmop{new}, j 2})$ and set it to be the
same as $\nu_j$. Suppose the selected one is $\nu_{\tmop{new}, j 1}$. 
We then draw $\nu_{\tmop{new},j 2}$ 
from the set of neighboring states of $\nu_{j}$, denoted as $\mathcal{N}_{\nu_{j}}$. 
Instead of proposing randomly from $\mathcal{N}_{\nu_{j }}$, we use a carefully specified informed proposal distribution function, $g_{\tmop{informed}} (\nu_{\tmop{new}, j 2} \mid
\nu_{j}, \mathcal{D}_{\tmop{new},j 2}^{\ast})$, where $\mathcal{D}_{\tmop{new},j 2}^{\ast}$ in $g_{\tmop{informed}}(\cdot \mid \cdot)$ indicates that it uses the corresponding local information for proposal. $g_{\tmop{informed}} (\nu_{\tmop{new}, j 2} \mid
\nu_{j}, \mathcal{D}_{\tmop{new},j 2}^{\ast})$ favors neighboring states with higher posterior probabilities gains while satisfying local detailed balancing conditions~\citep{zanella2020informed}. See  Supplementary material for more details of $g_{\tmop{informed}}(\cdot \mid \cdot)$.
One can use parallel and Cholesky updating/downdating data likelihood recursive algorithms to reduce the computations required for evaluating $g_{\tmop{informed}}(\cdot\mid \cdot)$ for all possible neighboring states. 
The acceptance ratio for the birth move is computed as
\begin{align*}
  & \min \left\{ 1, \frac{\lambda}{k_{\tmop{new}}} \cdot \frac{r_d (k_{\tmop{new}})}{r_b (k)}
  \cdot \frac{\mathbb{P} (\nu_{\tmop{new}, j 1} \mid \mathcal{D}_{\tmop{new},
  j 1}^{\ast}) \mathbb{P} (\nu_{\tmop{new}, j 2} \mid \mathcal{D}_{\tmop{new},
  j 2}^{\ast})}{\mathbb{P} (\nu_j \mid \mathcal{D}_j^{\ast})} \cdot
  \frac{\mathbb{P}(\tmmathbf{y} \mid \pi_{\tmop{new}}^{\ast},
   k_{\tmop{new}}, \mathcal{T}_{\tmop{new}}, \tmmathbf{\nu}_{\tmop{new}}, \mathcal{X},
  \mathcal{S})}{\mathbb{P}(\tmmathbf{y} \mid \pi^{\ast},  k,
  \mathcal{T},\tmmathbf{\nu}, \mathcal{X}, \mathcal{S})} \right.  \\
  & \quad\quad\quad\quad\quad\quad\left. \cdot \frac{1}{g_{\tmop{informed}} (\nu_{\tmop{new}, j 2} \mid
  \nu_{j}, \mathcal{D}_{\tmop{new}, j 2}^{\ast})} \right\}, & 
\end{align*}
where 
$k_{\tmop{new}}=k+1,\mathcal{T}_{\tmop{new}}=\mathcal{T}$, $\mathbb{P}(\nu\mid \mathcal{D}_j^\ast):=\mathbb{P}(\nu\mid \|\mathcal{D}_j^\ast\|)$, and $\|\mathcal{D}_j^\ast\|$ is the number of locations within domain $\mathcal{D}_j^\ast$.

For the death move, we randomly merge two adjacent clusters in $\pi^{\ast}$.
Specifically, an edge in $\mathcal{T}$ that connects two distinct clusters in
$\pi^{\ast}$ is selected uniformly, then the two clusters are combined into a
single cluster. Let $\mathcal{D}^{\ast}_{\tmop{new}, j j'}
=\mathcal{D}_j^{\ast} \cup \mathcal{D}_{j'}^{\ast}$ be the new cluster merged
from $\mathcal{D}_j^{\ast}$ and $\mathcal{D}_{j'}^{\ast}$. We assign
$\nu_{\tmop{new}, j j'}$, the active set of $\mathcal{D}^{\ast}_{\tmop{new}, j
j'}$, as one of  ($\nu_j$, $\nu_{j'}$)  with an equal probability. 

For the change move, we first perform a death move, then a birth move to
update $\pi^{\ast}$. The purpose of the change move is to encourage better mixing
of the sampler. The cluster number is unchanged after the change move. Suppose
clusters $\mathcal{D}_j^{\ast}$ and $\mathcal{D}_{j'}^{\ast}$ are first merged
into one cluster $\mathcal{D}^{\ast}_{\tmop{new}, j j'}$ in the death move. For
the subsequent birth move, we split $\mathcal{D}^{\ast}_{\tmop{new}, j j'}$
into two clusters, say $\mathcal{D}_{\tmop{new},j }^{\ast}$ and
$\mathcal{D}_{\tmop{new},j' }^{\ast}$. Let $\nu_{\tmop{new},j }$ and $\nu_{\tmop{new},j'}$ be the corresponding new active sets. We set $(\nu_{\tmop{new},j}, \nu_{\tmop{new},j'})$ to be a random permutation of $(\nu_j,
\nu_{j'})$. See Section Supplementary Material for acceptance ratios derived for death and change moves.

Finally, for the hyper move, we update the spanning tree $\mathcal{T}$ compatible with the current partition following a similar MST algorithm as in \cite{luo2021bayesian}. The active sets $\tmmathbf{\nu}$ are unchanged in the hyper move, and the acceptance ratio is $1$. 

In practice, to facilitate mixing, we also update $\tmmathbf{\nu}\mid  \pi^{\ast}, k, \mathcal{T}, \mathcal{X}, \mathcal{S}, \tmmathbf{y}$ after the RJ-MCMC move. This update can be performed using either a standard MCMC step in the Bayesian variable selection context \citep{yang2016computational}, or the informed proposal $g_{\tmop{informed}}(\cdot \mid \cdot)$.


After collecting the posterior samples of
$(\pi^{\ast}, \tmmathbf{\nu}, \tmmathbf{\theta})$, we can predict
$\mu (\tmmathbf{s})$ at a new observed location $\tmmathbf{s}$ by $\mu
(\tmmathbf{s}) = \tmmathbf{x}^T_{\nu (\tmmathbf{s})} (\tmmathbf{s})
\tmmathbf{\theta} (\tmmathbf{s}) $, where $\tmmathbf{\theta} (\tmmathbf{s}) =
\sum_{j = 1}^k \tmmathbf{\theta}_j \mathbb{I} (\tmmathbf{s} \in
\mathcal{D}^{\ast}_j)$ and $\nu (\tmmathbf{s}) = \sum_{j = 1}^k \nu_j
\mathbb{I} (\tmmathbf{s} \in \mathcal{D}^{\ast}_j)$ are the predicted
regression coefficient and active set, respectively.
\section{Theoretical results}\label{SEC:theory}

In this section, we introduce theoretical results under the VS
setting. 
We assume that there is a true domain partition of $\mathcal{D}$, say
$\{ \mathcal{D}_{l, 0} \}_{l = 1}^{k_0}$, such that the true regression mean
\begin{equation}
    \mu_0 (\tmmathbf{s}) = \tmmathbf{x}^T(\tmmathbf{s})\tilde{\tmmathbf{\theta}}_0(\tmmathbf{s}),\text{ where }\tilde{\tmmathbf{\theta}}_0 (\tmmathbf{s}) = \sum_{l = 1}^{k_0}
\mathbb{I} (\tmmathbf{s} \in \mathcal{D}_{l, 0})
\tilde{\tmmathbf{\theta}}_{l, 0},\label{EQ:reformulation}
\end{equation}
$k_0$ is the true number of clusters, and $\tilde{\tmmathbf{\theta}}_{l, 0}$ is the true regression coefficient vector with sparsity at $\mathcal{D}_{l,0}$. Let $\nu_{l,0}\subseteq \{1,\ldots,q\}$ be the true active set corresponding to $\tilde{\tmmathbf{\theta}}_{l, 0}$. We
will establish the posterior consistency theory for $\{ \mathcal{D}_{l, 0} \}_{l = 1}^{k_0},
\{\nu_{l,0}\}_{l=1}^{k_0}, \{\tilde{\tmmathbf{\theta}}_{l,0}\}_{l=1}^{k_0}$ and study the prediction
error. We consider a high-dimensional context, where $q$ may go to $\infty$
and satisfy $n^{- \tau} \log (q) \rightarrow 0$ for some $\tau < 1$. We place particular emphasis on studying the effects of prior hyperparameter choices on posterior contraction properties.



\subsection{Notations}

Denote $\| \cdummy \|_2$ and $\| \cdummy \|_{\infty}$ as $L_2$ norm and
$L_{\infty}$ norm, respectively. The notation $|\mathbb{D}|$ is used as two
ways: if $\mathbb{D}$ is a spatial domain, $|\mathbb{D}|$ represents its area;
if $\mathbb{D}$ is a set, $|\mathbb{D}|$ denotes its cardinality.

For two locations $\tmmathbf{s}_1, \tmmathbf{s}_2$, let $d
(\tmmathbf{s}_1, \tmmathbf{s}_2) = \|\tmmathbf{s}_1 -\tmmathbf{s}_2 \|_2$ be
the Euclidean distance. For two \ spatial domains $\mathbb{D}_1$ and
$\mathbb{D}_2$, we write $d (\mathbb{D}_1, \mathbb{D}_2) =
\inf_{\tmmathbf{s}_1 \in \mathbb{D}_1, \tmmathbf{s}_2 \in \mathbb{D}_2} d
(\tmmathbf{s}_1, \tmmathbf{s}_2)$. For a constant $\delta > 0$ and a spatial
domain $\mathbb{D}$, we define the $\delta$-neighborhood of $\mathbb{D}$ as
$ \mathcal{N} (\mathbb{D}, \delta) = \{\tmmathbf{s} \in \mathbb{R}^2 : d
   (\tmmathbf{s}, \mathbb{D}) \leq \delta\}$.

We use $c$ and $C$ to denote some constants independent of 
$\pi^{\ast}$ and $\tmmathbf{\nu}$. The values of $c$ and $C$ may change from
line to line. For two positive series $a_n$ and $b_n$, we say $a_n \gg b_n$ if $a_n / b_n
\rightarrow \infty$, and $a_n \ll b_n$ if $a_n / b_n \rightarrow 0$. We write
$a_n \sim b_n$ if there exist constants $c, C > 0$, such that $c < a_n / b_n <
C, \text{for all } n \geqslant 1$. We write $a_n = O (b_n)$, if there exits a
constant $c > 0$, such that $a_n / b_n \leqslant c$ for all $n \geqslant 1$.
We write $a_n = o (b_n)$ if $a_n \ll b_n$.

For a matrix $\tmmathbf{A}$ and a constant $c$, we say $\tmmathbf{A}> c$, if
$\lambda_{\min} (\tmmathbf{A}) > c$, and $\tmmathbf{A}< c$ if $\lambda_{\max}
(\tmmathbf{A}) < c$, where $\lambda_{\min} (\tmmathbf{A})$ and $\lambda_{\max}
(\tmmathbf{A})$ are the minimum and maximum eigenvalues of $\tmmathbf{A}$,
respectively.

\subsection{Assumptions and main theorems}

Before presenting our main theorems, we first introduce the necessary assumptions. 

{\assumption{\label{AS:distributionloc}
$\{\tmmathbf{s}_i \}_{i = 1}^n$ are independent and identically distributed
($i.i.d.$) on spatial domain $\mathcal{D}$ with a probability density function
$\mathbb{P}_{\mathcal{D}} (\cdummy)$ satisfying $0 < \inf_{\tmmathbf{s} \in
\mathcal{D}} \mathbb{P}_{\mathcal{D}} (\tmmathbf{s}) \leqslant
\sup_{\tmmathbf{s} \in \mathcal{D}} \mathbb{P}_{\mathcal{D}} (\tmmathbf{s}) <
\infty$.}}

We next define the boundary set of the true partition $\{\mathcal{D}_{l, 0} \}_{l = 1}^{k_0}$ as
$ \mathcal{B}= \{\tmmathbf{s} \in \mathcal{D}: \text{there exist } l \neq l',
   \text{such that } \mathcal{D}_{l, 0} \cap \mathcal{N}(\tmmathbf{s}, \delta)
   \neq \emptyset, \text{and } \mathcal{D}_{l', 0} \cap
   \mathcal{N}(\tmmathbf{s}, \delta) \neq \emptyset \text{ hold for }\forall
   \delta > 0\}$. 
{\assumption{\label{AS:lengthofboundary}We assume that $\mathcal{B}$ and
$\{\mathcal{D}_{l, 0} \}_{l = 1}^{k_0}$ satisfy
\begin{enumerate}[label=\theassumption.\arabic*,ref=\theassumption.\arabic*]
  \item \label{AS:lengthofboundary1}The boundary set $\mathcal{B}$ has a
  $\delta$-covering number $N (\mathcal{B}, \delta, \| \cdot \|_2) \leqslant c
  \delta^{- 1}$ for some constant $c > 0$.
  
  \item \label{AS:lengthofboundary2}For a given $l \in \{1, \ldots, k_0 \}$
  and any two locations $\tmmathbf{s}, \tmmathbf{s}' \in \mathcal{D}_{l, 0}$,
  there exists a path connecting $\tmmathbf{s}$ and $\tmmathbf{s}'$, say
  $\mathcal{P} (\tmmathbf{s}, \tmmathbf{s}')$, such that $\mathcal{P}
  (\tmmathbf{s}, \tmmathbf{s}')$ is contained in $\mathcal{D}_{l, 0}$ and
  \begin{equation}
    \label{EQ:pathdis} d \{\mathcal{P}(\tmmathbf{s}, \tmmathbf{s}'),
    \mathcal{B}\} \geqslant \min \{d (\tmmathbf{s}, \mathcal{B}), d
    (\tmmathbf{s}', \mathcal{B}), C\},
  \end{equation}
  where $C > 0$ is some constant.
\end{enumerate}}}

Assumption \ref{AS:distributionloc} is standard in the spatial statistics literature \citep{luo2021bayesian,yu2024distributed}. Assumption \ref{AS:lengthofboundary} assumes mild conditions on the shape of $\{\mathcal{D}_{l,0}\}_{l=1}^{k_0}$. Specifically, Assumption
\ref{AS:lengthofboundary1} is satisfied if the boundary set
$\mathcal{B}$ is a curve with finite length \citep{luo2021bayesian}. Assumption \ref{AS:lengthofboundary2}
assumes each sub-domain $\mathcal{D}_{l, 0}$ to be connected, and it can be shown that (\ref{EQ:pathdis}) is satisfied for some common shapes (e.g., finite union of regular
polygons and circles). Under Assumption \ref{AS:lengthofboundary}, we have the following proposition from \cite{huangandsang2025consistency}. 

\begin{proposition}
  \label{PP:bestapproximation}Under Assumption \ref{AS:lengthofboundary},
  there exists a contiguous domain partition $\pi^{\ast}_0 = \{\mathcal{D}_{1,
  0}^{\ast}, \ldots \mathcal{D}_{k_0, 0}^{\ast} \}$ in our partition model
  space, such that
  \begin{equation}
    |\mathcal{W}_{\pi_0^{\ast}} | = \sum_{j = 1}^{k_0} |\{B_m : B_m \subseteq
    \mathcal{D}_{j, 0}^{\ast}, B_m \subsetneq \mathcal{D}_{j, 0} \}| \leqslant
    cK \label{EQ:bestpai}
  \end{equation}
  for some constant $c$, where $\mathcal{W}_{\pi_0^{\ast}} = \cup_{j =
  1}^{k_0} \{B_m : B_m \subseteq \mathcal{D}_{j, 0}^{\ast}, B_m \subsetneq
  \mathcal{D}_{j, 0} \}$ is the set of blocks containing mis-clustered
  locations.
\end{proposition}

Proposition~\ref{PP:bestapproximation} establishes the existence of a partition $\pi^{\ast}_0$ within the model's partition space such that the number of blocks containing mis-clustered locations, denoted by $\mathcal{W}_{\pi_0^{\ast}}$, is bounded by $cK$. In what follows, we define the ``approximation error" as the area of $\mathcal{W}_{\pi_0^{\ast}}$. Since each block has area $K^{-2}$, Proposition~\ref{PP:bestapproximation} implies that the approximation error is bounded by $K^{-2}\times cK = O(K^{-1})$. Thus, increasing $K$ reduces the approximation error.


{\assumption{\label{AS:thetagap}There exists a positive constant $c$, such
that $\min_{l \neq l'} \| \tilde{\tmmathbf{\theta}}_{l, 0} -
\tilde{\tmmathbf{\theta}}_{l', 0} \|_2 > c > 0$.}}

{\assumption{\label{AS:covariates}Conditional on $\mathcal{S}$, $\{
\tmmathbf{x}_{\dagger} (\tmmathbf{s}_i) \}_{i = 1}^n$ are independent random variables.
We assume $0 < c \leqslant \mathbb{E} [\{ \tmmathbf{x}_{\nu} (\tmmathbf{s})
\tmmathbf{x}_{\nu}^T (\tmmathbf{s}) \} \mid \tmmathbf{s}] \leqslant C$ and $\|
\tmmathbf{x} (\tmmathbf{s}) \|_{\infty} \leqslant C$ for all
$\tmmathbf{s} \in \mathcal{D}$, and $\nu \in \mathcal{I}$ with $| \nu |
\leqslant 3q_{\max}$.}}

Assumption \ref{AS:thetagap} ensures sufficient separation between clusters, which is necessary for identifying the sub-domains $\{\mathcal{D}_{l, 0}\}_{l = 1}^{k_0}$. We consider a random covariate design, where Assumption \ref{AS:covariates} serves to prevent collinearity among covariates. See also Assumption (A2) in \cite{yu2024distributed} and Condition (C1) in {\cite{luo2021bayesian}} for similar assumptions. 

Let $r_1, r_2 > 0$ be two positive values. The next assumption is on the orders of 
hyperparameters, which provide practical guidance for specifying priors in our model. 

{\assumption{\label{AS:hyperpara}We assume that the
number of blocks hyperparameter $K$, variable number penalty function $\alpha
(\cdot)$ in (\ref{EQ:priornu}), and the Poisson hyperparameter $\lambda$ in
(\ref{DEF:k}) satisfy
\begin{enumerate}[label=\theassumption.\arabic*,ref=\theassumption.\arabic*]
  \item \label{AS:K} $\sqrt{\frac{n}{\log
  (q + 1) \log^{1 + r_1} (n)}} \gg K \gg \log^{1 + r_1 + r_2} (n)$
  
  \item \label{AS:nu} $\alpha (\cdot)$ is a non-decreasing function with first
  derivative continuous and $\lim_{x \rightarrow
  \infty} \alpha' (x) \rightarrow 0$. For any given constant $c$, we assume $n \log^{- r_2} (n) \gg \alpha (cn) \gg \{ n
  K^{- 1} + \log (q + 1) \} \log^{1 + r_1} (n)$ 
  
  
  \item \label{AS:lambda}$\lambda = o (1)$, with $n \gg \log (\lambda^{- 1})
  \gg \max \{ \alpha (n), K^2 \log^{r_1} (n) \}$
\end{enumerate}}}

We take a moment to explain the rationale behind Assumptions \ref{AS:K} - \ref{AS:lambda}. Following Proposition \ref{PP:bestapproximation}, we demonstrate that a larger $K$ reduces the approximation error of $\pi^\ast_0$. However, as the number of possible block partitions grows exponentially with $K^2$, a larger $K$ increases the complexity of identifying the correct partition. Thus, a trade-off is required in selecting $K$. 
Besides, note that a larger $q$ corresponds to a larger active set space $\mathcal{I}$, which further increases the difficulty of identifying the correct partition. Assumption \ref{AS:K} accounts for this by requiring a smaller upper bound of $K$ as $q$ increases, and ensures a proper balance between the complexity of the joint space of partitions and active sets, and the approximation error.

Following (\ref{EQ:priornu}), $\alpha(\cdot)$ is evaluated at the sample size (i.e., the number of locations) within each cluster. The non-decreasing condition on $\alpha(\cdot)$ in Assumption \ref{AS:nu} aligns with the classical VS principle that the penalty increases with sample size \citep{zhao2006model}. The derivative condition on $\alpha(\cdot)$ is imposed for technical reasons. A simple example satisfying $\lim_{x \rightarrow \infty} \alpha'(x) \rightarrow 0$ is a function independent of $x$, for which $\alpha'(x) \equiv 0$. 
A trade-off governs the choice of $\alpha(\cdot)$: it must be sufficiently large to prevent selecting extraneous variables, yet not excessively large to exclude variables in the true active set. Note that the lower bound in Assumption \ref{AS:nu} depends on $nK^{-1}$ and the number of variables $q$. By Assumption \ref{AS:distributionloc} and Proposition \ref{PP:bestapproximation}, $nK^{-1}$ is approximately the number of locations in $\mathcal{W}_{\pi_0^{\ast}}$. Meanwhile, as $q$ increases, the risk of overfitting grows due to the larger active set space $\mathcal{I}$. Thus, the lower bound in Assumption \ref{AS:nu} ensures that both the locations in $\mathcal{W}_{\pi_0^{\ast}}$ and the effect of $q$ on overfitting are negligible compared with $\alpha(cn)$.

Another trade-off arises in determining the rate of $\lambda$. Since the model with a larger number of clusters offers enhanced flexibility
for data fitting, the data likelihood tends to favor a larger number of
clusters. To avoid such overfitting, the Poisson hyperparameter $\lambda$
serves as a penalty for the number of clusters, yielding the condition
$\lambda = o (1)$ in Assumption \ref{AS:lambda}. There are three considerations to guide the rate choice of $\lambda$. First, the decay rate of  $\lambda$ must not be too fast; otherwise, the number of clusters may be underestimated. Second, to enhance data fitting (i.e., increase the likelihood), the model can either introduce more clusters or select more variables within each cluster. The condition $\log (\lambda^{- 1}) \gg \alpha (n)$ ensures that the penalty for the cluster number dominates that on the number of variables, thereby encouraging the model to prioritize selecting more variables to improve the fit before increasing the number of clusters. Third, a smaller $\lambda$ should be used for a larger partition space (i.e., a larger $K$), since it is more likely to overfit. These considerations collectively yield the rate condition specified in Assumption \ref{AS:lambda}.

In Remark \ref{RM:hyper}, we provide an example of hyperparameters satisfying Assumption \ref{AS:hyperpara}. The next assumption is on the priors of our model. Recall that we use $\Delta$ to denote the space of spanning trees induced from $\mathcal{G}$.

{\assumption{\label{AS:prior}We make the following assumptions on the priors
of our model.
\begin{enumerate}[label=\theassumption.\arabic*,ref=\theassumption.\arabic*]
  \item \label{AS:k} The true number of clusters $k_0$ satisfies $k_0 \leq
  k_{\max}$, for $k_{\max}$ specified in (\ref{DEF:k}).
  
  \item \label{AS:spanningtree} For $\pi_0^{\ast}$ in Proposition
  \ref{PP:bestapproximation}, we assume
  \[ \sup_{\mathcal{T}_1 \in \Delta, \mathcal{T}_2 \in \{\mathcal{T} \in
     \Delta \text{: } \pi_0^{\ast}  \text{ can be induced from } \mathcal{T}\}}
     \left\{ \frac{\mathbb{P}(\mathcal{T}_1)}{\mathbb{P}(\mathcal{T}_2)}
     \right\} = O [\exp \{cK \log (K)\}] \]
  for some constant $c$.
  
  \item \label{AS:qmax} We assume $\max_{1 \leqslant l \leqslant k_0} | \nu_{l, 0} | \leqslant
  q_{\max}$.
\end{enumerate}}}

Assumptions \ref{AS:k} and \ref{AS:qmax} ensure the true number of clusters and active sets are within our
prior. Assumption \ref{AS:spanningtree} assumes that the spanning tree's probability of inducing $\pi^\ast_0$ is not excessively small. For the UST prior specified in Section~\ref{subsec:spanningtree},
Assumption \ref{AS:spanningtree} is satisfied immediately since the prior
ratio of any two spanning trees is $1$.

We next introduce a distance
measure between two spatial domain partitions. Let $\pi_1 (\mathcal{D}) =
\{\mathcal{D}_{11}, \ldots, \mathcal{D}_{1 k_1} \}$ and $\pi_2 (\mathcal{D}) =
\{\mathcal{D}_{21}, \ldots, \mathcal{D}_{2 k_2} \}$ be two domain partitions, where $k_1$ and $k_2$ are
their respective number of clusters. We write
\begin{equation}
  \epsilon \{\pi_1 (\mathcal{D}), \pi_2 (\mathcal{D})\} = 2 - |\mathcal{D}|^{-
  1} \Bigl[ \sum_{j = 1}^{k_1} \max_{l \in \{1, \ldots, k_2 \}}
  |\mathcal{D}_{1 j} \cap \mathcal{D}_{2 l} | + \sum_{l = 1}^{k_2} \max_{j \in
  \{1, \ldots, k_1 \}} |\mathcal{D}_{1 j} \cap \mathcal{D}_{2 l} | \Bigr] 
  \label{DEF:epsilonspatial}
\end{equation}
as their distance measure. Similarly, for two partitions of $\mathcal{S}$, say $\pi_1 (\mathcal{S}) =
\{\mathcal{S}_{11}, \ldots, \mathcal{S}_{1 k_1} \}$ and $\pi_2 (\mathcal{S}) =
\{\mathcal{S}_{21}, \ldots, \mathcal{S}_{2 k_2} \}$, where $k_1$ and $k_2$ are their respective number of clusters, the distance
between $\pi_1 (\mathcal{S})$ and $\pi_2 (\mathcal{S})$ is defined as
\begin{equation}
  \epsilon_n \{\pi_1 (\mathcal{S}), \pi_2 (\mathcal{S})\} = 2 - n^{- 1} \Bigl[
  \sum_{j = 1}^{k_1} \max_{l \in \{1, \ldots, k_2 \}} |\mathcal{S}_{1 j} \cap
  \mathcal{S}_{2 l} | + \sum_{l = 1}^{k_2} \max_{j \in \{1, \ldots, k_1 \}}
  |\mathcal{S}_{1 j} \cap \mathcal{S}_{2 l} | \Bigr] .
  \label{DEF:epsilonnlocation}
\end{equation}
Our partition consistency theory is established based on $\epsilon (\cdot, \cdot)$ and $\epsilon_n (\cdot, \cdot)$. Similar distance measures were adopted in \cite{van2000performance} and \cite{huangandsang2025consistency}.
Write $\mathfrak{D}= \{ \mathcal{X}, \mathcal{S}, \tmmathbf{y} \}$ and $\{\mathcal{S}_{l, 0} =
\{ \tmmathbf{s}_i \in \mathcal{D}_{l, 0}, 1 \leqslant i \leqslant n \}\}_{l=1}^{k_0}$ as the true partition of $\mathcal{S}$. Let $0 <
r_0 \leqslant 1 + r_1$ be a constant. We have the following result.

\begin{theorem}
  \label{TH:clustererror}Under Assumptions \ref{AS:distributionloc},
  \ref{AS:lengthofboundary}, \ref{AS:thetagap}, \ref{AS:covariates},
  \ref{AS:hyperpara} and \ref{AS:prior}, there exists a positive constant $c$, such that with probability tending to $1$, we have
  \begin{equation}
    \mathbb{P} (| \pi^{\ast} | = k_0 \mid \mathfrak{D}) \geqslant 1 - \exp \{-
    c n K^{- 1} \log^{r_0} (n) \}, \label{EQ:kconsistency}
  \end{equation}
  \begin{equation}
    \mathbb{P} (\epsilon [\pi^{\ast}, \{ \mathcal{D}_{l, 0} \}_{l = 1}^{k_0}]
    \leqslant cK^{- 1} \log^{r_0} (n) \mid \mathfrak{D}) \geqslant 1 - \exp
    \{- c n K^{- 1} \log^{r_0} (n) \},\text{ and} \label{EQ:errorate0}
  \end{equation}
  \begin{equation}
    \mathbb{P} (\epsilon_n [\pi, \{ \mathcal{S}_{l, 0} \}_{l = 1}^{k_0}]
    \leqslant c K^{- 1} \log^{r_0} (n) \mid \mathfrak{D}) \geqslant 1 - \exp
    \{- c n K^{- 1} \log^{r_0} (n) \} . \label{EQ:errorrate}
  \end{equation}
\end{theorem}

Under Assumption \ref{AS:hyperpara}, the right hand sides of Equations (\ref{EQ:kconsistency}) - (\ref{EQ:errorrate}) converge to $1$, and $K^{-1}\log^{r_0}(n) \rightarrow 0$. Thus, Theorem \ref{TH:clustererror} establishes the partition consistency of our model: with probability tending to $1$, the posterior partition recovers the correct number of clusters, and the posterior distributions of $\pi^\ast$ and $\pi$ achieve a contraction rate of $K^{-1}\log^{r_0}(n)$ around the truth.

\begin{remark}\label{RM:hyper}
Note that the contraction rate  $K^{-1}\log^{r_0}(n)$ decreases as $K$ increases. Thus, a larger $K$ is preferred to achieve a faster contraction rate. From the upper bound in Assumption \ref{AS:K}, the maximal admissible order of $K$ is $K\sim \sqrt{\frac{n}{\log (q+1) \log^{1 + r_b} (n)}}$ for some $r_b>0$. With this choice of $K$, we can further set $\alpha(x)=c_{a} x/\log^{r_a}(1+x)$ for some $c_a,r_a>0$, and $\log (\lambda^{- 1}) \sim n \log^{- r_p} (n)$, for some $0 <r_p < \min\{1+r_b,r_a\}$. It can be verified that Assumption \ref{AS:hyperpara} is satisfied under these hyperparameter settings.
\end{remark}

For a given $\pi^{\ast} = \{\mathcal{D}_1^{\ast},
\ldots, \mathcal{D}^{\ast}_k \}$, we write $\mathcal{M} (\mathcal{D}^{\ast}_j) =
\tmop{argmax}_{l \in \{1, \ldots, k_0 \}} |\mathcal{D}^{\ast}_j \cap
\mathcal{D}^{\ast}_{l, 0} |$ as the index of the sub-domain in $\{
\mathcal{D}^{\ast}_{l, 0} \}_{l = 1}^{k_0}$ with the largest intersection area with
$\mathcal{D}^{\ast}_j$. $\mathcal{D}^{\ast}_{\mathcal{M}
(\mathcal{D}^{\ast}_j), 0}$ is considered as the ``best matched" sub-domain in $\{
\mathcal{D}^{\ast}_{l, 0} \}_{l = 1}^{k_0}$ for  $\mathcal{D}^{\ast}_j$. Thus, roughly speaking, we consider $\nu_{\mathcal{M}(\mathcal{D}^{\ast}_j), 0}$ as the ``true" active set at $\mathcal{D}^{\ast}_j$. Let $\nu_0 (\tmmathbf{s}) = \sum_{l = 1}^{k_0}
\mathbb{I} (\tmmathbf{s} \in \mathcal{D}_{l, 0}) \nu_{l, 0}$ be the true
active set at $\tmmathbf{s}$ and recall its prediction $\nu
(\tmmathbf{s})$ in Section \ref{SEC:computation}. The next theorem establishes posterior consistency for active sets. 

\begin{theorem}
  \label{TH:selectionconsistency}Under Assumptions \ref{AS:distributionloc},
  \ref{AS:lengthofboundary}, \ref{AS:thetagap}, \ref{AS:covariates},
  \ref{AS:hyperpara} and \ref{AS:prior}, with probability tending to 1, we
  have
  \begin{equation}
    \mathbb{P} [\{ \nu_{\mathcal{M} (\mathcal{D}_j^{\ast}), 0} \}_{j = 1}^k =
    \{ \nu_{l, 0} \}_{j = 1}^{k_0} \mid \mathfrak{D}] \rightarrow 1,
    \label{EQ:TH21}
  \end{equation}
  \begin{equation}
    \mathbb{P} (\nu_j = \nu_{\mathcal{M} (\mathcal{D}_j^{\ast}), 0}, \forall j
    \mid \mathfrak{D}) \rightarrow 1, \text{ and} \label{EQ:TH22}
  \end{equation}
  \begin{equation}
    \mathbb{P} \Bigl[ \int \mathbb{P}_{\mathcal{D}} (\tmmathbf{s})\mathbb{I}
    \{ \nu (\tmmathbf{s}) \neq \nu_0 (\tmmathbf{s}) \} d \tmmathbf{s} > M_n
    K^{- 1} \log^{r_0} (n) \mid \mathfrak{D} \Bigr] \rightarrow 0
    \label{EQ:TH23}
  \end{equation}
  for any sequence $M_n \rightarrow \infty$.
\end{theorem}

Equations (\ref{EQ:TH21}) - (\ref{EQ:TH22}) show that with probability tending to 1, the set $\{\nu_{\mathcal{M}(\mathcal{D}_j^\ast), 0} \}_{j = 1}^k$ is the same as the set $\{\nu_{l, 0} \}_{l = 1}^{k_0}$, and the posterior of $\nu_j$ concentrates on its ``true" value $\nu_{\mathcal{M} (\mathcal{D}_j^{\ast}), 0}$. Equation (\ref{EQ:TH23}) establishes that the area where the active set is mis-predicted vanishes at a rate of $K^{-1}\log^{r_0}(n)$.

Given $(\pi^{\ast}, \tmmathbf{\nu}, \tmmathbf{\theta})$, write $\tilde{\tmmathbf{\theta}}_j$ as the sparse regression coefficient induced from $(\nu_j,\tmmathbf{\theta}_j)$ at $\mathcal{D}_{j}^{\ast}$, and write $\tilde{\tmmathbf{\theta}} (\tmmathbf{s}) = \sum_{j
= 1}^k \tilde{\tmmathbf{\theta}}_j \mathbb{I} (\tmmathbf{s} \in
\mathcal{D}^{\ast}_j)$ as the prediction of $\tilde{\tmmathbf{\theta}}_0(\tmmathbf{s})$. Recall the regression mean prediction
$\mu (\tmmathbf{s})$ in Section \ref{SEC:computation}. We have the following result.
\begin{theorem}
  \label{TH:thetaconsistency}Under Assumptions \ref{AS:distributionloc},
  \ref{AS:lengthofboundary}, \ref{AS:thetagap}, \ref{AS:covariates},
  \ref{AS:hyperpara} and \ref{AS:prior}, with probability tending to $1$, we
  have
  \begin{equation}
    \mathbb{P} [\{
    \tilde{\tmmathbf{\theta}}_{\mathcal{M}(\mathcal{D}^{\ast}_j), 0} \}_{j =
    1}^k =\{ \tilde{\tmmathbf{\theta}}_{l, 0} \}_{l = 1}^{k_0} \mid
    \mathfrak{D}] \rightarrow 1, \label{EQ:TH31}
  \end{equation}
  \begin{equation}
    \mathbb{P} \{\max_{1 \leqslant j \leqslant k} \|
    \tilde{\tmmathbf{\theta}}_j -
    \tilde{\tmmathbf{\theta}}_{\mathcal{M}(\mathcal{D}^{\ast}_j), 0} \|_2 >
    M_n' K^{- 1} \log^{r_0} (n) \mid \mathfrak{D}\} \rightarrow 0,
    \label{EQ:TH32}
  \end{equation}
  \begin{equation}
    \mathbb{P} \Bigl\{ \int \mathbb{P}_{\mathcal{D}}
    (\tmmathbf{s}) \| \tilde{\tmmathbf{\theta}} (\tmmathbf{s}) -
    \tilde{\tmmathbf{\theta}}_0 (\tmmathbf{s}) \|_2^2  d \tmmathbf{s} > M_n'' K^{- 1} \log^{r_0} (n) \mid
    \mathfrak{D} \Bigr\} \rightarrow 0,\text{ and} \label{EQ:TH33}
  \end{equation}
  \begin{equation}
    \mathbb{P} \Bigl[ \int \mathbb{P}_{\mathcal{D}} (\tmmathbf{s}) \{ \mu
    (\tmmathbf{s}) - \mu_0 (\tmmathbf{s}) \}^2 d\tmmathbf{s}> M_n''' K^{- 1}
    \log^{r_0} (n) \mid \mathfrak{D} \Bigr] \rightarrow 0 \label{EQ:TH34}
  \end{equation}
  for any sequences $M'_n, M_n'', M_n''' \rightarrow \infty$.
\end{theorem}

Similar to the interpretation of $\nu_{\mathcal{M}(\mathcal{D}^{\ast}_j), 0}$, we consider $\tilde{\tmmathbf{\theta}}_{\mathcal{M}(\mathcal{D}^{\ast}_j), 0}$ as the ``true" regression coefficient at $\mathcal{D}_{j}^{\ast}$. Thus, $\|\tilde{\tmmathbf{\theta}}_j -\tilde{\tmmathbf{\theta}}_{\mathcal{M}(\mathcal{D}^{\ast}_{j}), 0}\|_2$ in (\ref{EQ:TH32}) is considered as the distance between
$\tilde{\tmmathbf{\theta}}_j$ and its ``true" value. Equations (\ref{EQ:TH31}) - (\ref{EQ:TH32}) show that with probability tending to 1, the set $\{\tilde{\tmmathbf{\theta}}_{\mathcal{M}(\mathcal{D}_j^\ast), 0} \}_{j = 1}^k$ is the same as the set $\{\tilde{\tmmathbf{\theta}}_{l, 0} \}_{l = 1}^{k_0}$, and the posterior contraction rate of $\tilde{\tmmathbf{\theta}}_j$ is of the order $K^{-1}\log^{r_0}(n)$. Equations
(\ref{EQ:TH33}) - (\ref{EQ:TH34}) provide the {{posterior predictive contraction rates}} of $\tilde{\tmmathbf{\theta}}(\tmmathbf{s})$ and $\mu(\tmmathbf{s})$, respectively.

\begin{remark}\label{RM:commonrate}
    In our hierarchical priors, $\tmmathbf{\nu}$ and $\tilde{\tmmathbf{\theta}}(\tmmathbf{s})$ are obtained conditional on $\pi^\ast$, implying that the contraction rates associated with $\tmmathbf{\nu}$ (Equation (\ref{EQ:TH23})) and $\tilde{\tmmathbf{\theta}}(\tmmathbf{s})$ (Equation (\ref{EQ:TH33})) are typically no faster than that of $\pi^\ast$. Moreover, note that the contraction rates provided in Equations (\ref{EQ:TH23}) and (\ref{EQ:TH33}) coincide with the contraction rate of $\pi^\ast$, i.e., $K^{-1}\log^{r_0}(n)$, in Equation (\ref{EQ:errorate0}). This suggests that the partition estimation error dominates the subsequent parameter estimation errors.
\end{remark}

\begin{remark}
    Under the hyperparameter setting in Remark \ref{RM:hyper} and following Remark \ref{RM:commonrate}, the posterior contraction rate of our model is $K^{- 1} \log^{r_0} (n) \sim n^{-1/2}\log^{1/2}(q)\log^{r_0+(1+r_b)/2}(n)$, which decays more slowly with a larger $q$. This aligns with intuition, as a larger $q$ increases the model complexity, making consistency more difficult to achieve.
\end{remark}

\section{Simulation}\label{SEC:simu}

\begin{figure}[h]
		\centering
		\begin{tabular}{cccc}
             {\includegraphics[width=0.22\linewidth,height=0.20\textheight]{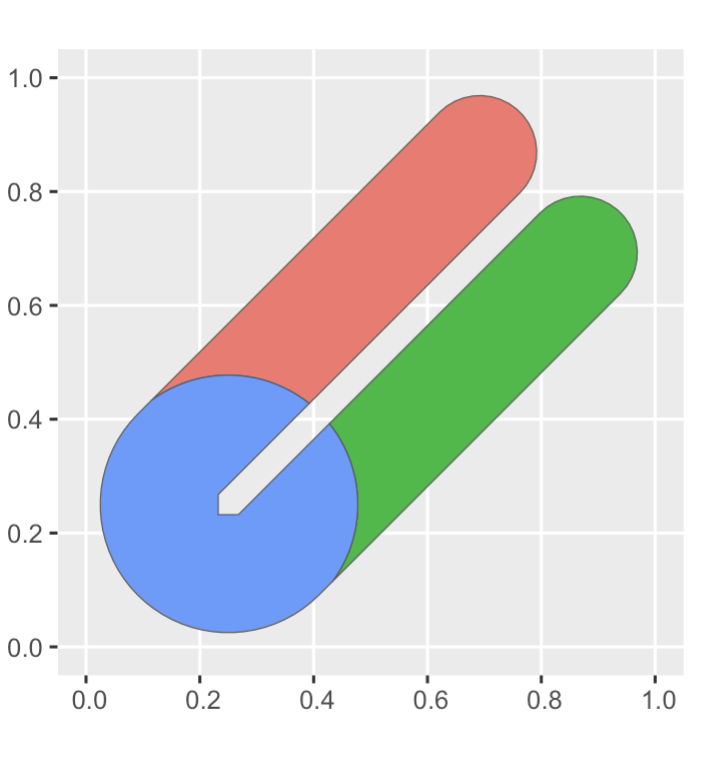}}&
			{\includegraphics[width=0.25\linewidth,height=0.205\textheight]{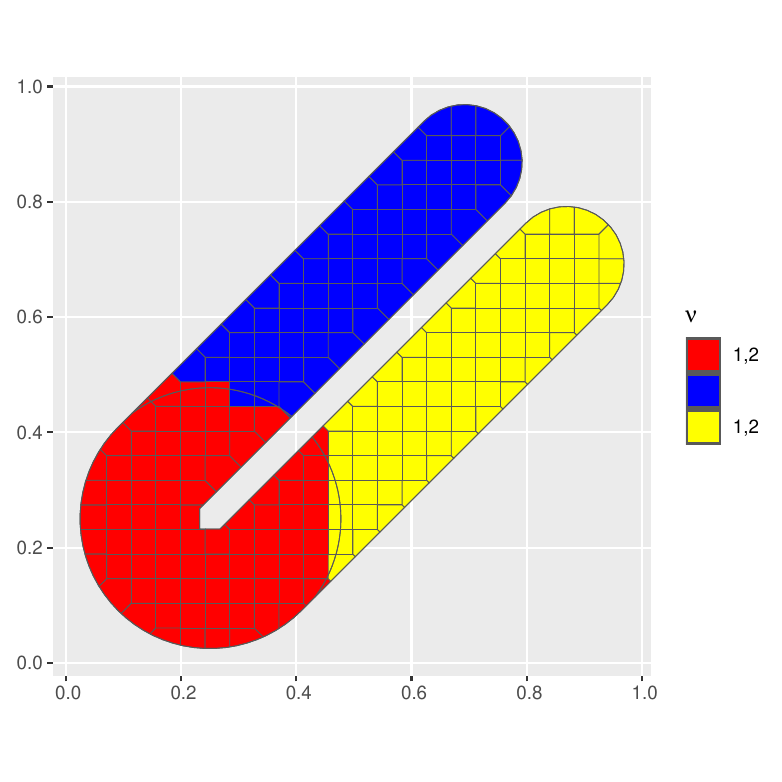}}
            &{\includegraphics[width=0.25\linewidth,height=0.205\textheight]{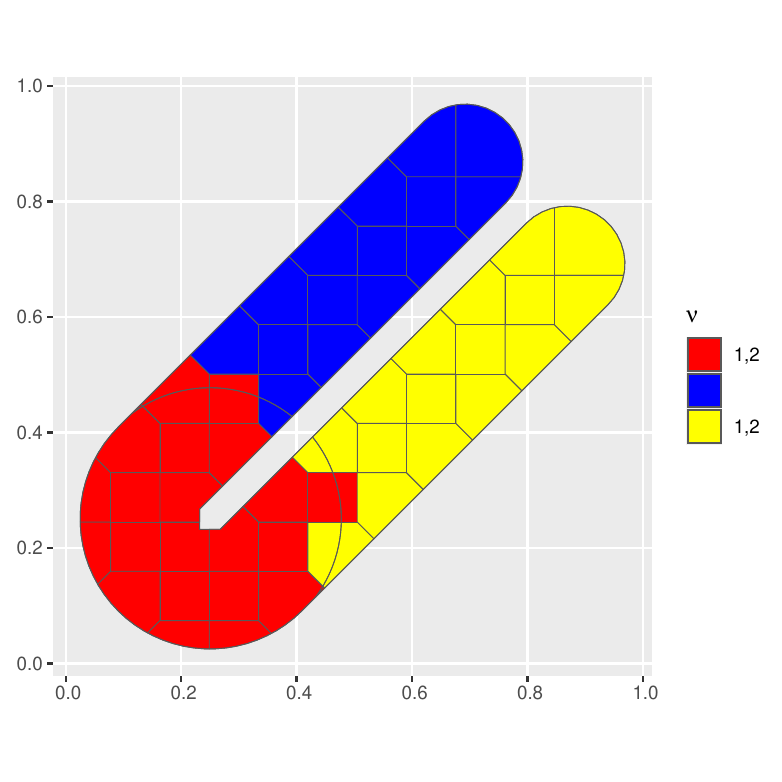}}
            &{\includegraphics[width=0.22\linewidth,height=0.2\textheight]{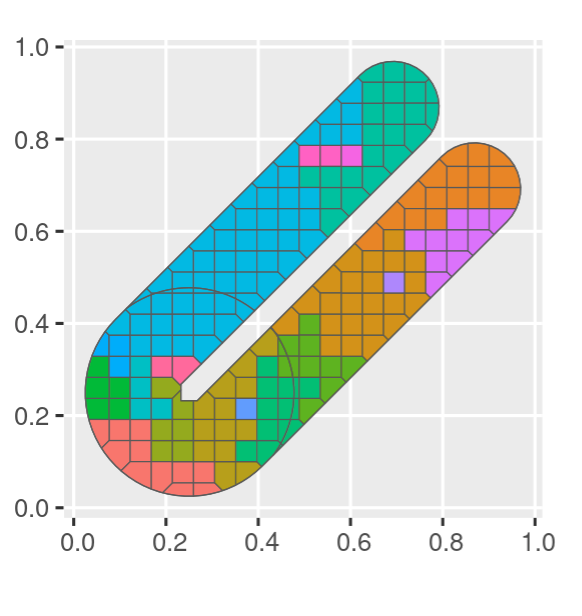}}\\
			{\small (a)} & {\small (b)} & {\small (c)}  & {\small (d)} 
		\end{tabular}
		\caption{(a) U-shape domain with $\{\mathcal{D}_{l,0}\}_{l=1}^{3}$ represented by different colors. (b-d) One randomly chosen posterior partition sample with different colors representing different clusters, under the VS, SVS and NR settings, respectively. Legends $\nu$ in (b-c) show the corresponding posterior active sets for different clusters. Note that the goal of NR is adaptive prediction, not partition recovery. 
        }
		\label{FIG:n3000result}
        \vspace{-10pt}
	\end{figure}

\noindent\textbf{Data generation.} We conduct simulations under the VS, SVS, and NR settings. 
In all three settings, we generate $n$ locations from a U-shape domain with three clusters as shown in Figure \ref{FIG:n3000result}(a), where $\mathcal{D}_{1,0}$ and $\mathcal{D}_{2,0}$ are the upper and lower arms, and $\mathcal{D}_{3,0}$ is the middle circle. 
Under the VS and SVS settings, we generate $q$-dimensional covariates by $\tmmathbf{x}_{\dagger}(\tmmathbf{s}_i) = \mathbf{R}_X^T \tmmathbf{z}(\tmmathbf{s}_i)$, where $\{\tmmathbf{z}(\tmmathbf{s}_i)\}_{i=1}^{n}$ are $i.i.d.$ $\tmop{Unif}[-1,1]$ random variables and $\mathbf{R}_X$ is set to be the Cholesky factor of the $q\times q$ matrix $\{3\exp(-|p_1-p_2|)\}_{p_1=1,p_2=1}^{q,q}$. 
The local active sets are $\nu_{1,0}=\emptyset$, $\nu_{2,0}=\nu_{3,0}=\{1,2\}$.  
    In VS, the cluster-wise regression coefficients for $\tmmathbf{x}^T(\cdot)=\{1,\tmmathbf{x}^T_{\dagger}(\cdot)\}$ are $\tilde{\tmmathbf{\theta}}^T_{1,0}=(1,\tmmathbf{0}^T)$, $\tilde{\tmmathbf{\theta}}^T_{2,0}=(1,1,-1,\tmmathbf{0}^T)$ and $\tilde{\tmmathbf{\theta}}^T_{3,0}=(1,-1,1,\tmmathbf{0}^T)$, where $\tmmathbf{0}$ denotes a zero vector whose dimension may vary depending on the context. In SVS, we adopt the same cluster-wise regression coefficients for $\tmmathbf{x}^T_{\dagger}(\cdot)$ as VS, but replace the spatially clustered intercept with a spatially smoothly varying intercept modeled by $f(\tmmathbf{s})=\sin\{2\pi(s_1+s_2)\}$.  
In NR, we generate a regression mean function of a 2-D coordinates, $(s_1,s_2)$, assuming $\mu_{1,0}(\tmmathbf{s})=s_1s_2$, $\mu_{2,0}(\tmmathbf{s})=\sin\{2\pi(s_1+s_2)\}$, and $\mu_{3,0}(\tmmathbf{s})=\sum_{k=1}^{50}0.5^{k}\cos[3^{k}\pi\{s_1 \cos(\pi/4)+s_2 \sin(\pi/4)\}]$, where $\mu_{l,0}(\cdot)$ is the regression mean function at $\mathcal{D}_{l,0},1\leq l \leq 3$. We set $\{\epsilon(\tmmathbf{s}_i)\}_{i=1}^{n}$ to be $i.i.d.$ Gaussian noises with standard deviation $1$ in VS and NR. In SVS, we reduce the noise standard deviation to $0.1$ to mitigate the effect of functional approximation bias induced by the nonparametric component.


\noindent\textbf{Model settings.}
We fit the VS, SVS, and NR models under different simulation settings, respectively. 
We run $5000$ MCMC iterations with a burn-in period $4000$ and a thinning parameter $5$. 
In VS, to investigate the asymptotic property of our model, we experiment with $n \in \{100, 500, 1000, 2000, 3000\}$ and $q=\lfloor\exp(n^{0.2})\rfloor$, where $\lfloor a \rfloor$ denotes the greatest integer less than or equal to $a$.  Following Remark \ref{RM:hyper} in the theoretical results, we set
	the number of blocks hyperparameter $K$, the variable number penalty function $\alpha(\cdot)$ and the Poisson hyperparameter $\lambda$ as $K = \lfloor c_b \sqrt{\frac{n}{\log (q+1) \log^{1 + r_b} (n)}}\rfloor$, $\alpha(x)=c_{a} x/\log^{r_a}(1+x)$, and $\log (\lambda^{- 1}) = c_p n \log^{- r_p} (n)$,  where $c_b=5, r_b = 1, c_a = 0.1, r_a = 0.5, c_p = 0.05$ and $r_p = 0.1$. 
We set $k_{\max} = 5$, $q_{\tmop{max}}=10$
	and $\sigma^2 = 1$. For each $n$, we repeat the simulation $100$ times. 

In SVS, we fix $n=3000$ and $q=\lfloor\exp(n^{0.2})\rfloor=142$.  We set $\tmmathbf{x}_0 (\cdot)$ as the bivariate quadratic spline bases \citep{lai2007spline}. 
To balance the increased model complexity due to the inclusion of nonparametric bases, we reduce the hyperparameter $K$ used in VS by 2 to shrink the partition space. Accordingly, we divide $\log(\lambda^{-1})$ by 2 to adjust the cluster number penalty in line with the reduced partition space. All other model settings remain the same as in VS. 

In NR, we fix $n = 3000$, and set $\tmmathbf{x}_0(\tmmathbf{s})\equiv 1$ and $\tmmathbf{x}_{\dagger}^T(\tmmathbf{s})=(s_1,s_2,s_1^2,s_1s_2,s_2^2)$.   Due to the approximation bias between $\tmmathbf{x}(\cdot)$ and $\{\mu_{l,0}(\cdot)\}_{l=1}^{3}$, the area of each cluster is required to shrink to $0$ as 
$n$ increases, ensuring that the approximation bias vanishes asymptotically. Consequently, the number of clusters diverges to infinity, 
so we set $k_{\tmop{max}}$ equal to the number of blocks. 
We fix $\alpha(x)\equiv 1$ and set $\sigma^2$ as the empirical variance of $\{y(\tmmathbf{s}_i)\}_{i=1}^{n}$. To select hyperparameters $(K,\lambda)$, we use Watanabe-Akaike information criterion (WAIC; {\citealp{watanabe2010asymptotic}}). The candidate values for $K$ are chosen such that each block contains $10,15$ or $20$ observed locations on average. The candidate values for $\lambda$ are chosen such that $\log(\lambda)\in\{5,6,7,8\}$.

\noindent\textbf{Competing methods.}
As reviewed in the Introduction, there is limited work on Bayesian spatial local variable selection in both the VS and SVS settings. In VS, we compare with the regularized spatially clustered coefficient (RSCC) model proposed in \cite{zhong2023sparse}. In SVS, 
among existing methods, the most relevant to our model is the spatial heterogeneous additive partial linear model (SHAPLM) proposed in \cite{Zhang17042025}. However, its computational burden becomes substantial when the number of covariates $q$ is large. When applied to our simulated dataset, SHAPLM encounters an out-of-memory error on a server with 100 GB RAM. In NR, we compare with the Bayesian Additive Regression Spanning Trees (BAST; \citealp{luo2021bast}), the bivariate spline model \citep{lai2007spline} fitted using the least squares method, and the nearest-neighbor Gaussian Process (GP) model in \cite{datta2016nearest}. 


\noindent\textbf{Results.}
Figures~\ref{FIG:n3000result}(b-d) show a randomly chosen posterior sample of the partition and active sets in VS, SVS and NR, respectively. Figures~\ref{FIG:n3000result}(b-c) show that our model recovers the active sets and closely matches the true domain partition in VS and SVS, except for some locations near boundaries, which is attributed to blocking approximation errors. 
Figure~\ref{FIG:n3000result}(d) shows the adaptivity of our model in NR: the middle circle region contains the most clusters because $\mu_{3,0}(\cdot)$ has the most irregular shape and therefore requires more clusters for accurate approximation. In contrast, the upper arm contains the fewest clusters, as $\mu_{1,0}(\cdot)$ can be exactly represented by bases in $\tmmathbf{x}(\cdot)$. However, in NR, there is no established theory guaranteeing partition consistency, as the diverging number of clusters required by our model does not converge to any fixed “true” cluster number. 

\begin{figure}[H]
		\centering
		\begin{tabular}{cccc}
             {\includegraphics[width=0.22\linewidth,height=0.2\textheight]{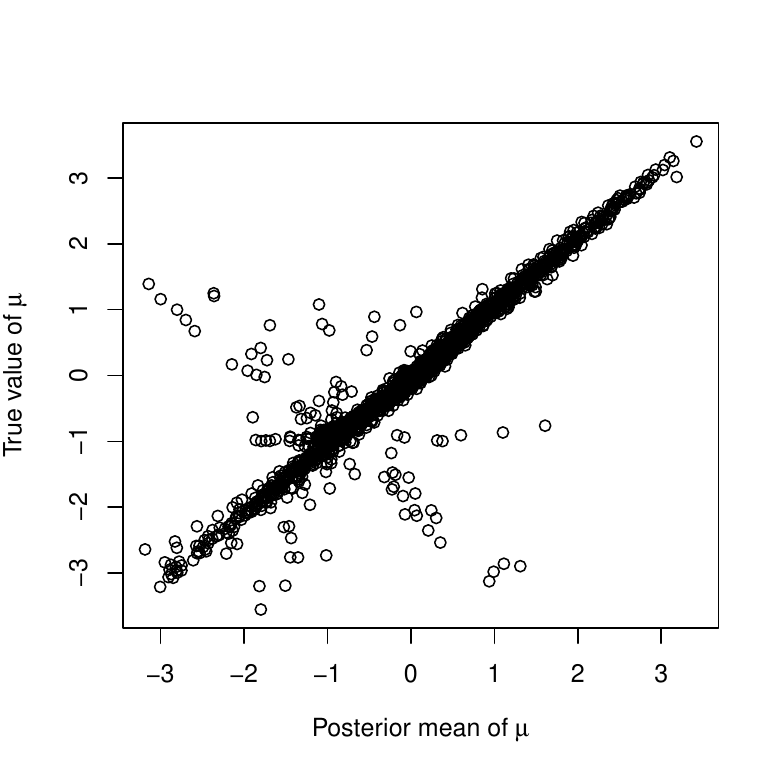}}&
			{\includegraphics[width=0.22\linewidth,height=0.25\textheight]{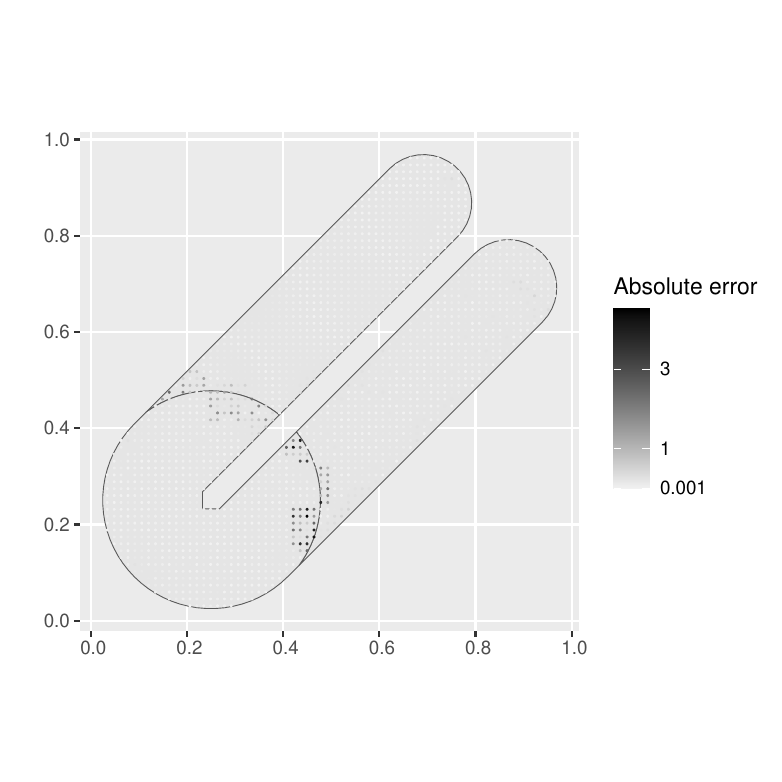}}
            &{\includegraphics[width=0.22\linewidth,height=0.2\textheight]{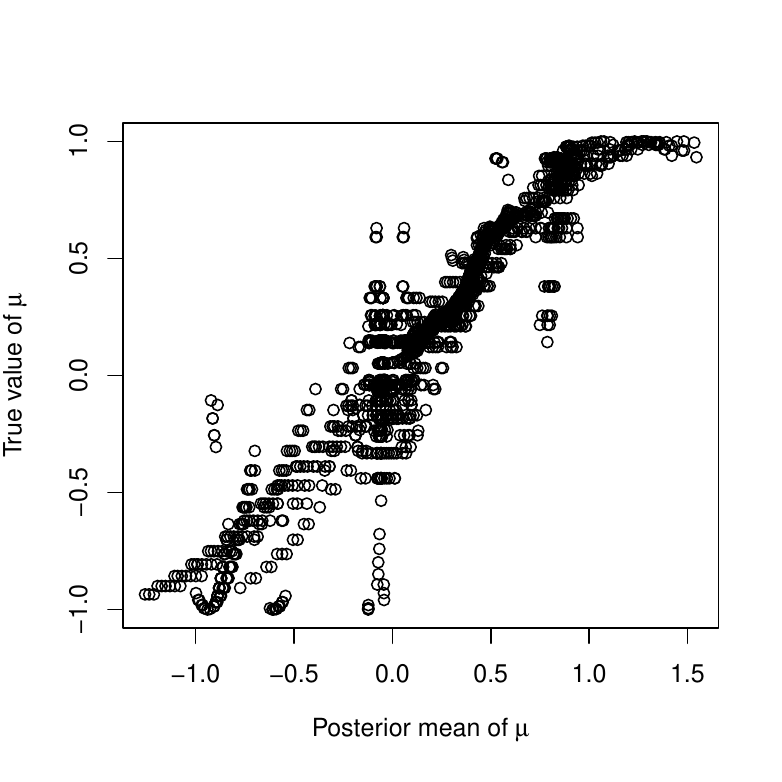}}
            &{\includegraphics[width=0.22\linewidth,height=0.25\textheight]{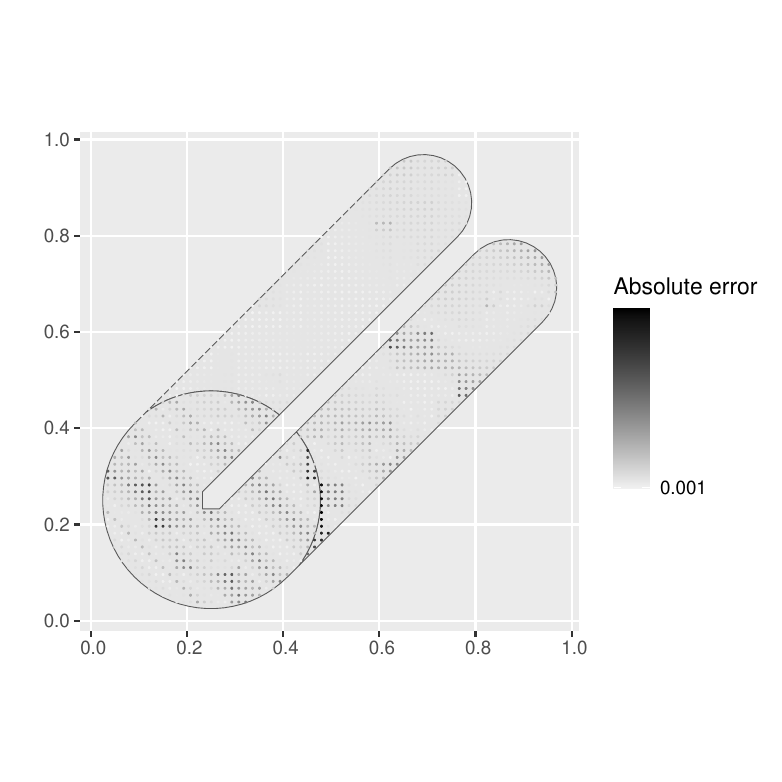}}\\
			{\small (a)} & {\small (b)} & {\small (c)}  & {\small (d)} 
		\end{tabular}
		\caption{(a) (c): Scatterplots of the posterior mean of $\mu(\tmmathbf{s})$ and the truth at $2000$ randomly distributed locations under SVS and NR, respectively. (b) (d): Spatial distribution of absolute prediction errors under SVS, NR setting, respectively. The darker the color, the larger the absolute prediction error.}
		\label{FIG:predresult}
        \vspace{-10pt}
	\end{figure}

Figure~\ref{FIG:predresult} presents the prediction results in SVS and NR. The prediction result of VS has a similar pattern to SVS and is thus omitted. Figures~\ref{FIG:predresult}(a) and (c) show that the posterior mean of $\mu(\tmmathbf{s})$ generally aligns with the truth, although some variability and bias are observed in NR. In Figure~\ref{FIG:predresult}(b), most prediction errors are small in SVS, except at misclustered boundary locations. Figure~\ref{FIG:predresult}(d) shows larger errors in the central circular region,  which is because $\mu_{3,0}(\cdot)$ has the most irregular shape.

\begin{figure}[h]
		\centering
		\begin{tabular}{ccc}
             {\includegraphics[width=0.31\linewidth,height=0.5\textheight]{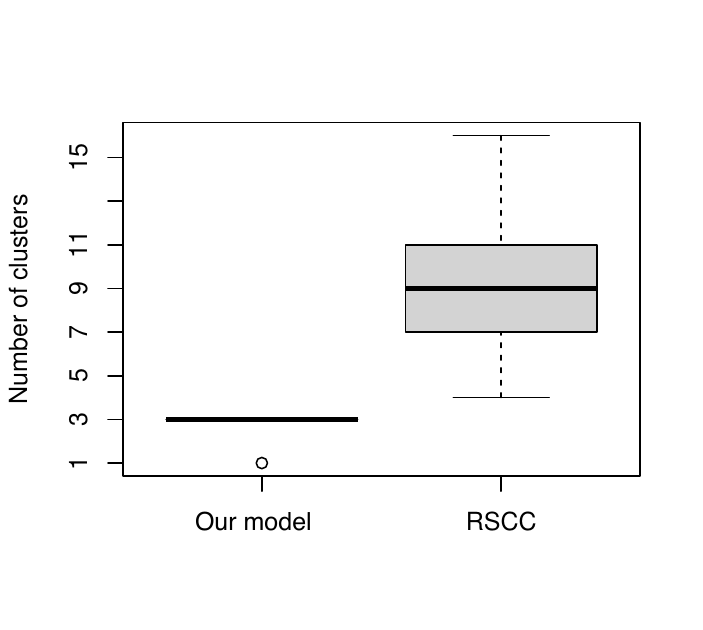}}
            &{\includegraphics[width=0.3\linewidth,height=0.2\textheight]{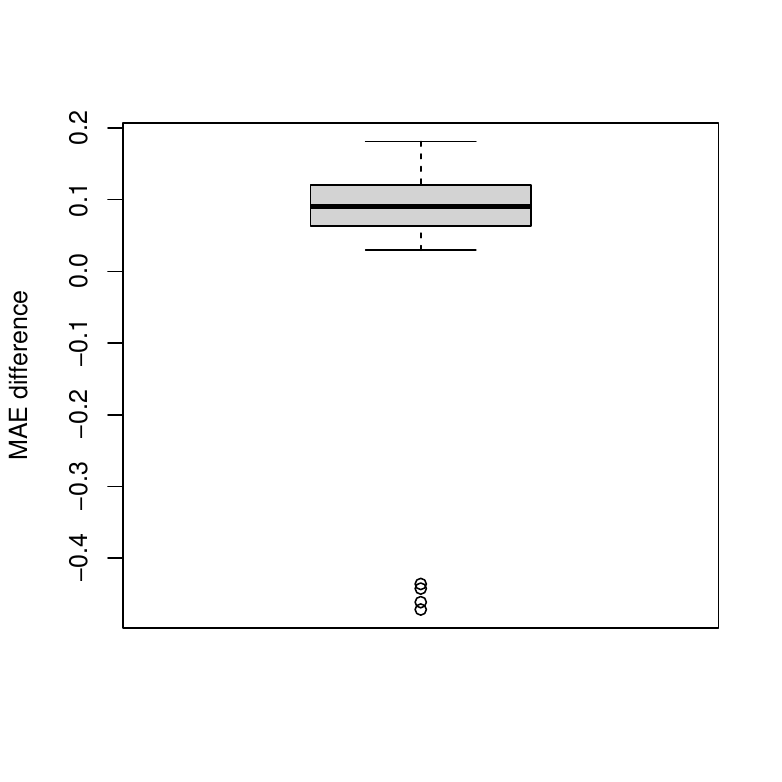}}
            &{\includegraphics[width=0.3\linewidth,height=0.25\textheight]{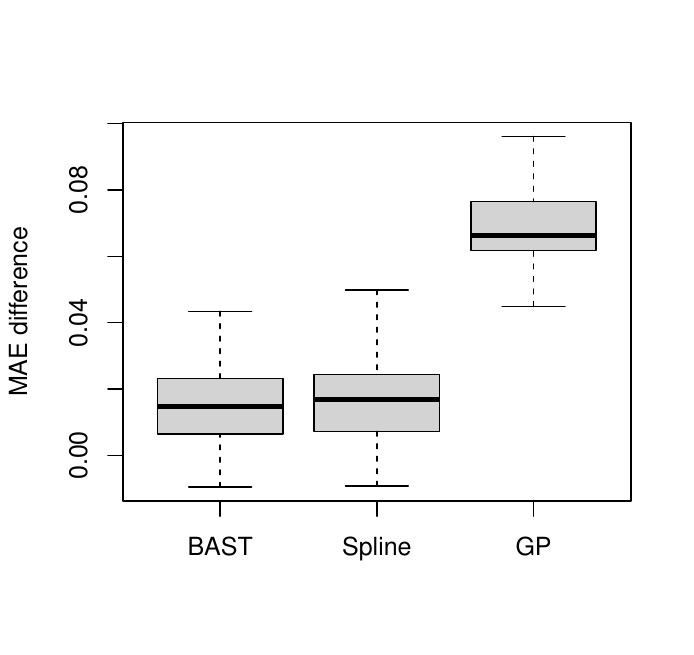}}\\
			{\small (a)} & {\small (b)} & {\small (c)}  
		\end{tabular}
		\caption{(a) Boxplots of the
			number of clusters over different repeats in VS. For each repeat, we take a random draw from our model's posterior cluster number samples.
			(b) Boxplot of MAE differences between RSCC and our model in VS. (c) Boxplots of MAE differences between each competing method and our model in NR.}
		\label{FIG:modelcompare}
        \vspace{-10pt}
	\end{figure}

Figures~\ref{FIG:modelcompare}(a-b) compare RSCC and our model in VS with $n=3000$: (a) shows the estimated number of clusters, and (b) shows the difference in mean absolute error (MAE) of the regression mean prediction. We can see that our model recovers the correct number of clusters in nearly all repeats, while RSCC tends to overestimate. 
Moreover, nearly all repeats yield a positive MAE difference, demonstrating that our model achieves higher prediction accuracy, likely due to the more accurate partition estimation.
Figure~\ref{FIG:modelcompare}(c) shows the MAE difference between each competing method and our model in NR. Most simulation repeats yield a positive MAE difference, indicating that our model outperforms competing models. This is attributed to the adaptivity of our model, enabled by spatial clustering and local basis selection. BAST approximates the regression mean function using piecewise constant functions, which is not optimal if the true mean function has higher-order derivatives. The spline model and GP adopt a global approach for function approximation and thus lack local adaptivity. Moreover, the standard GP relies on Euclidean distance, limiting its effectiveness in irregular domains such as the U-shape in this simulation.

Finally, we compare the computational time of RSCC and our model in VS. For a simulated data with $n = 3000$, we use the \texttt{microbenchmark} package in R to compute the average runtime over 5 repetitions. The result shows that our model and RSCC take 1279 and 2325 seconds on average, respectively. This demonstrates a substantial reduction in runtime, especially considering that our approach involves extensive sampling iterations, whereas RSCC relies on optimization. The improved efficiency is largely due to the reduced partition space, achieved through the blocking strategy and the informed MCMC technique described in Section~\ref{SEC:computation}.

\begin{figure}[hbtp]
\centering
\begin{tabular}{@{}c@{}c@{}c@{}c@{}c@{}c@{}}
\raisebox{0.12\textheight}{\small\text{(a)}} &
\begin{minipage}[t]{0.21\textwidth}
  \includegraphics[width=\linewidth,height=0.25\textheight]{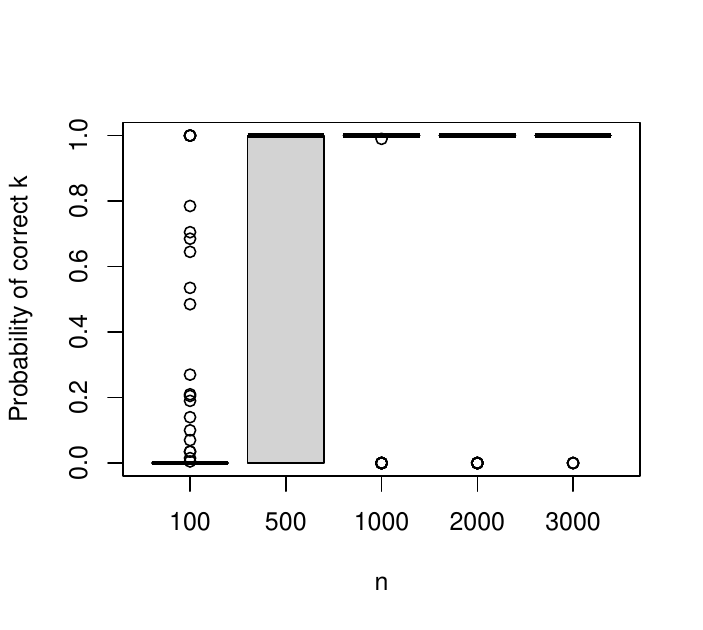}
\end{minipage} &
\raisebox{0.12\textheight}{\small\text{(c)}} &
\begin{minipage}[t]{0.21\textwidth}
  \includegraphics[width=\linewidth,height=0.25\textheight]{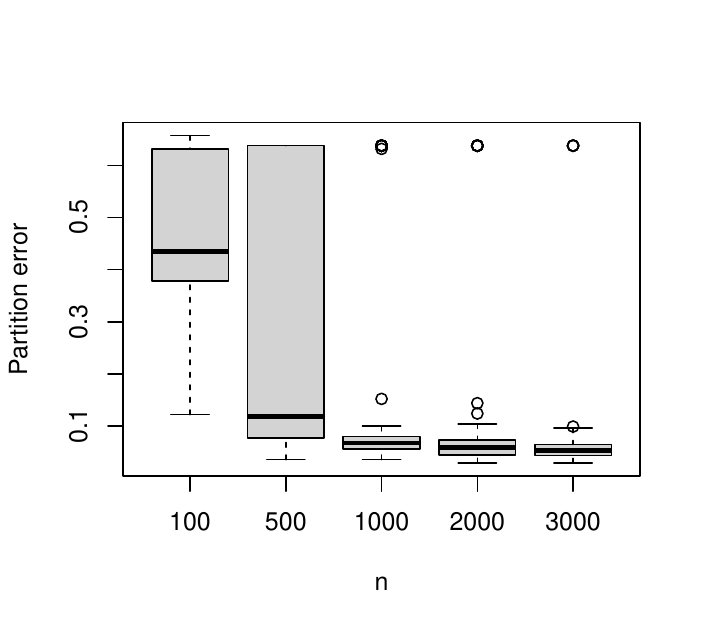}
\end{minipage}
\raisebox{0.12\textheight}{\small\text{(e)}} &
\begin{minipage}[t]{0.21\textwidth}
  \includegraphics[width=\linewidth,height=0.25\textheight]{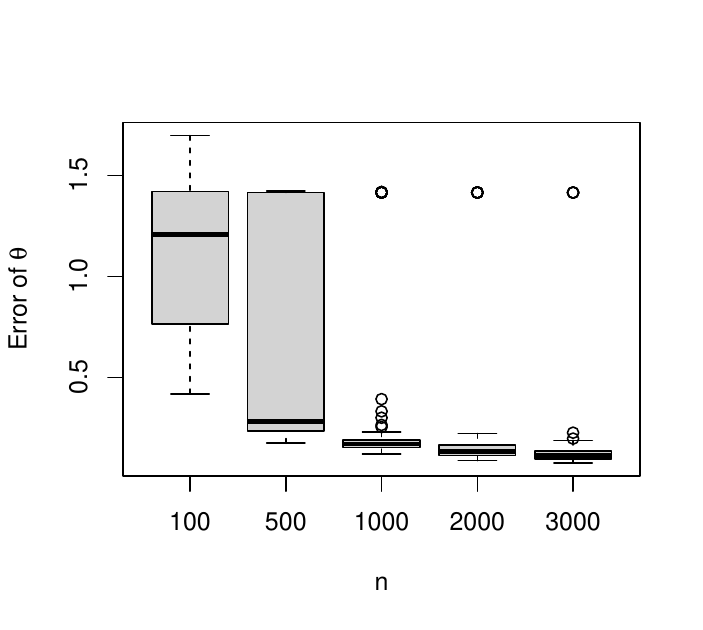}
\end{minipage}
\raisebox{0.12\textheight}{\small\text{(g)}} &
\begin{minipage}[t]{0.21\textwidth}
  \includegraphics[width=\linewidth,height=0.25\textheight]{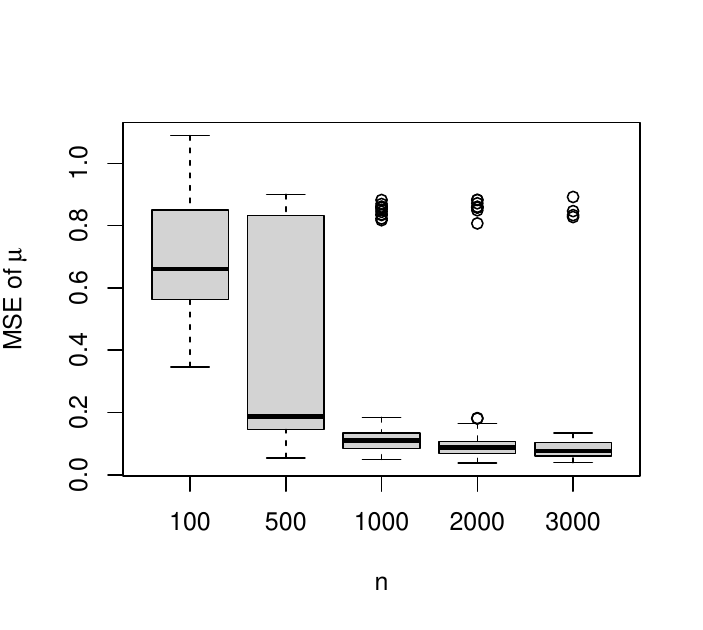}
\end{minipage}
\\
\raisebox{0.12\textheight}{\small\text{(b)}} &
\begin{minipage}[t]{0.21\textwidth}
  \includegraphics[width=\linewidth,height=0.25\textheight]{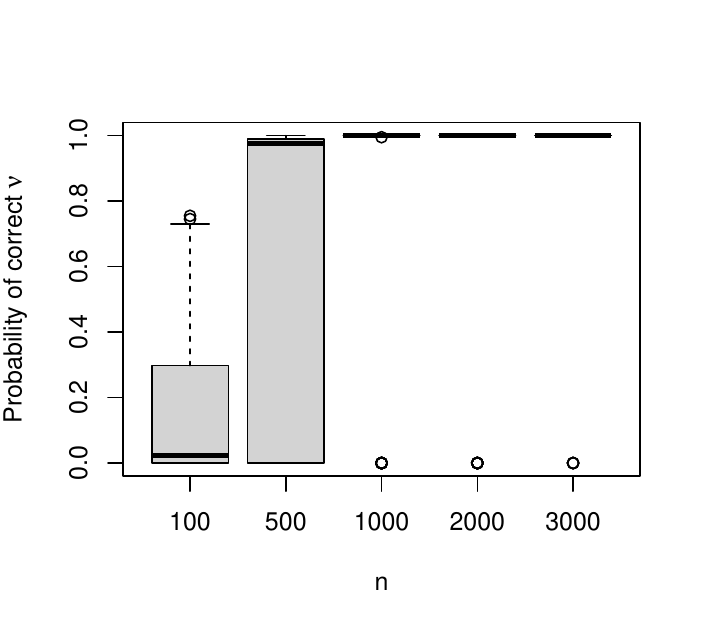}
\end{minipage} &
\raisebox{0.12\textheight}{\small\text{(d)}} &
\begin{minipage}[t]{0.21\textwidth}
  \includegraphics[width=\linewidth,height=0.25\textheight]{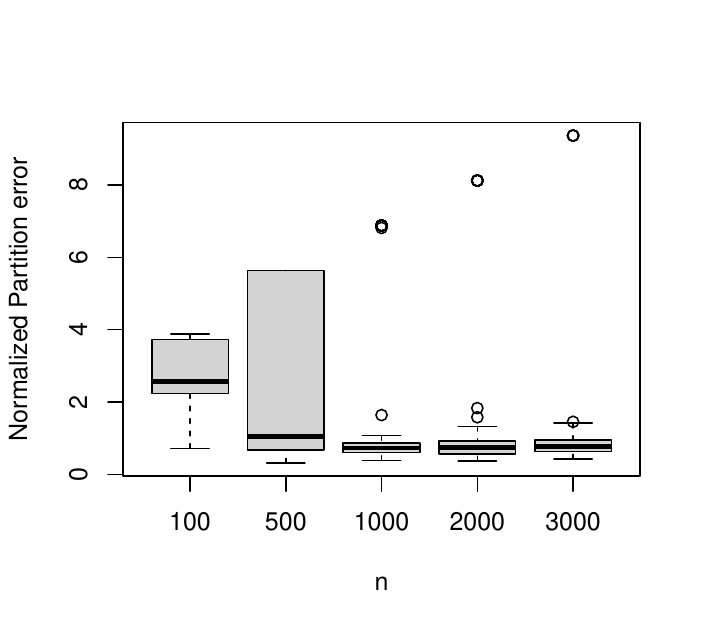}
\end{minipage}
\raisebox{0.12\textheight}{\small\text{(f)}} &
\begin{minipage}[t]{0.21\textwidth}
  \includegraphics[width=\linewidth,height=0.25\textheight]{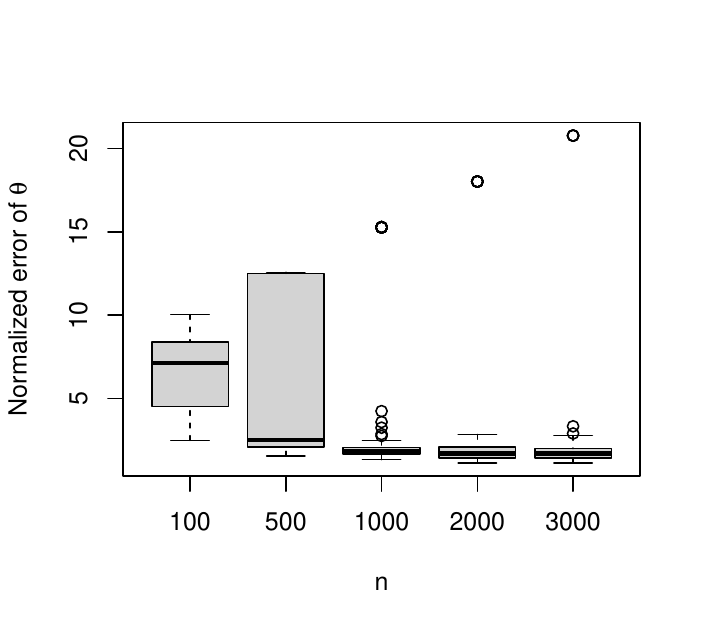}
\end{minipage}
\raisebox{0.12\textheight}{\small\text{(h)}} &
\begin{minipage}[t]{0.21\textwidth}
  \includegraphics[width=\linewidth,height=0.25\textheight]{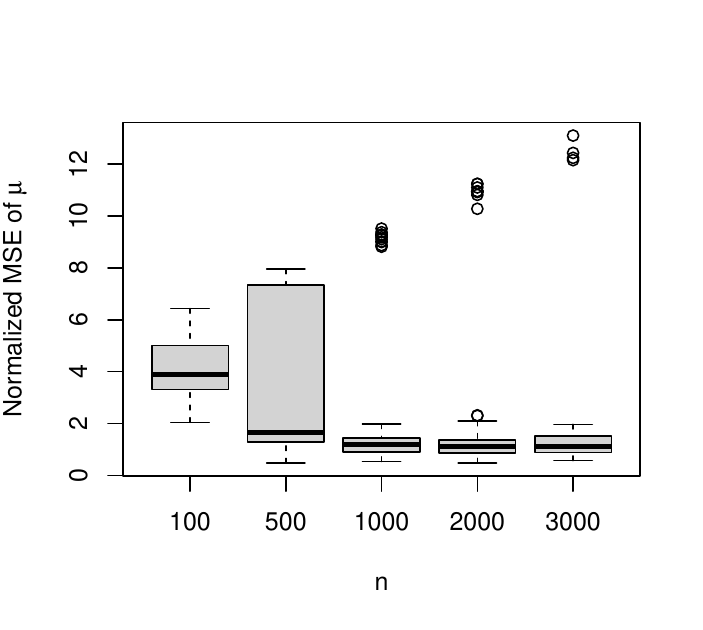}
\end{minipage}
\end{tabular}
		\caption{Fitting results under different values of $n$. (a) Boxplots of posterior probabilities of the true cluster number. (b) Boxplots of posterior probabilities of $\{\nu_{s,j} = \nu_{\mathcal{M}_{s,j},0},1\leq j \leq k_s\}$, where $ \mathcal{M}_{s, j}$ is the index of the sub-domain in $\{\mathcal{D}_{l,0}\}_{l=1}^{3}$ with the largest intersection area with the $j$-th cluster, at the $s$-th MCMC sample. (c) (e) (g) Boxplots of the posterior mean of $\epsilon[\pi^{\ast}_{s},\{\mathcal{D}_{l,0}\}_{l=1}^{3}]$, $\max_{1 \leqslant j \leqslant k_s} \|
    \tilde{\tmmathbf{\theta}}_{s,j} -
    \tilde{\tmmathbf{\theta}}_{\mathcal{M}_{s,j}, 0} \|_2 $, and $ m^{-1}\|\tmmathbf{\mu}_s-\tmmathbf{\mu}_0\|_2^2$, respectively. (d) (f) (h) Boxplots of the posterior mean of normalized counterparts of (c) (e) (g), respectively.}
		\label{FIG:symptotic_analysis}
\end{figure}

\noindent\textbf{Asymptotic analysis.}
Figure~\ref{FIG:symptotic_analysis} illustrates our model's performance under different values of $n$ in VS. As shown in Figures~\ref{FIG:symptotic_analysis}(a-b), when 
$n$ is small, the posterior number of clusters and active sets do not reliably recover the true values. However, after $n$ exceeds 1000, both quantities recover their true values with high probability, aligning with the consistency results in Equations~(\ref{EQ:kconsistency}) and~(\ref{EQ:TH22}).



Let $\{ k_s, \pi^\ast_s,\{\nu_{s,j}\}_{j = 1}^{k_s}
	\{ \tilde{\tmmathbf{\theta}}_{s, j} \}_{j = 1}^{k_s} \}_{s = 1}^M$
	denote the posterior samples of model parameters, where $k_s, \pi^\ast_s$, $\{\nu_{s,j}\}_{j = 1}^{k_s}$, $\{ \tilde{\tmmathbf{\theta}}_{s, j} \}_{j = 1}^{k_s}$  are the posterior cluster number, domain partition, active sets and regression
	coefficients, respectively, at the $s$-th MCMC sample.    
    Figures \ref{FIG:symptotic_analysis}(c) and (e) show posterior mean errors of $\pi^{\ast}_s$ and $\{ \tilde{\tmmathbf{\theta}}_{s, j} \}_{j = 1}^{k_s}$, respectively.  
    We can see the error decays as $n$ increases. According to Equations~(\ref{EQ:errorate0}) and~(\ref{EQ:TH32}), the theoretical contraction rate of the error is $K^{-1}\log^{r_0}(n)$, for a given $0<r_0\leq 1+r_1$. Setting $r_0=0.01$, we compute the ``normalized error" by rescaling the original error by $K^{-1}\log^{r_0}(n)$, and present the corresponding boxplots in Figures~\ref{FIG:symptotic_analysis}(d) and (f). We can see that when $n\geq 1000$, the distribution of normalized errors appears stable across different values of $n$, consistent with the theoretical result.

     To evaluate the regression mean prediction error of our model, we randomly generate $m=2000$ locations along with their covariates. 
     Let $\tmmathbf{\mu}_s$ denote the predicted regression mean vector at these sampled locations, given a posterior sample $[k_s, \pi^\ast_s,\{\nu_{s,j}\}_{j = 1}^{k_s}, \{ \tilde{\tmmathbf{\theta}}_{s, j} \}_{j = 1}^{k_s}]$.   Let $\tmmathbf{\mu}_0$ be the corresponding ground truth. Figures~\ref{FIG:symptotic_analysis}(g) and (h) show boxplots of the posterior mean of the prediction error $ m^{-1}\|\tmmathbf{\mu}_s-\tmmathbf{\mu}_0\|_2^2$ and its normalized counterpart, respectively. The result exhibits a pattern similar to that observed in Figures~\ref{FIG:symptotic_analysis}(c-f). Notably, for $n \geq 1000$, a few extreme errors appear in Figures~\ref{FIG:symptotic_analysis}(c-h), which are associated with incorrect partitioning, as indicated by outliers in Figure~\ref{FIG:symptotic_analysis}(a).

\section{Real data analysis} \label{SEC:realdata}


In this section, we analyze the SMO dataset introduced in Section~\ref{SEC:intro} using the proposed LSFS framework. Our goal is to identify heterogeneous tissue subregions, and within each subregion, select genes associated with CD8A protein expression. Identifying these genes offers insights into key transcriptional signals linked to CD8$^+$ T cell activity. By examining how different gene programs associate with CD8A across the tissue, we may potentially distinguish context-specific subtypes or functional states of CD8$^+$ T cells.

\begin{figure}[htbp]
\centering
\begin{tabular}{@{}c@{}c@{}c@{}c@{}c@{}c@{}}
\raisebox{0.12\textheight}{\small\text{(a)}} &
\begin{minipage}[t]{0.21\textwidth}
  \includegraphics[width=0.8\linewidth,height=0.18\textheight]{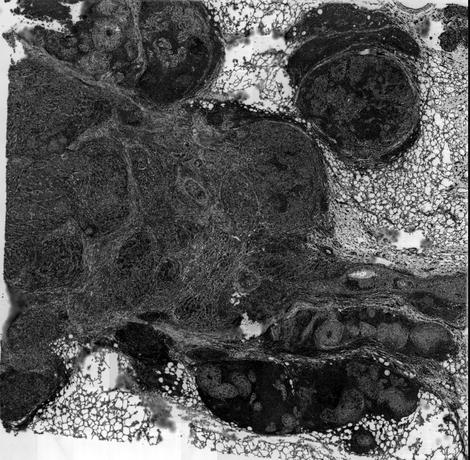}
\end{minipage} &
\raisebox{0.12\textheight}{\small\text{(b)}} &
\begin{minipage}[t]{0.21\textwidth}
  \includegraphics[width=1.15\linewidth,height=0.2\textheight]{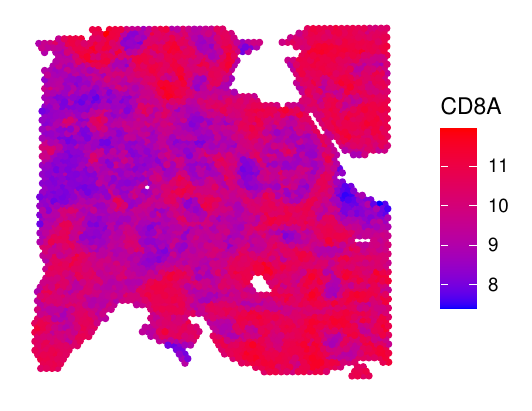}
\end{minipage}
\raisebox{0.12\textheight}{\small\text{(c)}} &
\begin{minipage}[t]{0.21\textwidth}
  \includegraphics[width=1.15\linewidth,height=0.2\textheight]{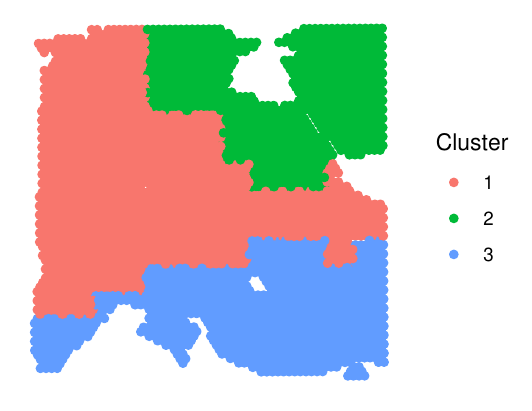}
\end{minipage}
\raisebox{0.12\textheight}{\small\text{(d)}} &
\begin{minipage}[t]{0.21\textwidth}
\hspace{0.05cm}
  \includegraphics[width=1.1\linewidth,height=0.135\textheight]{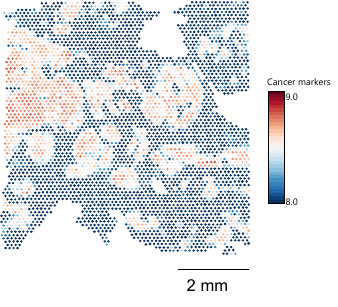}
\end{minipage}
\\
\raisebox{0.12\textheight}{\small\text{(e)}} &
\begin{minipage}[t]{0.21\textwidth}
  \includegraphics[width=\linewidth,height=0.195\textheight]{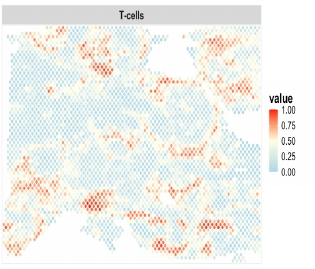}
\end{minipage} &
\raisebox{0.12\textheight}{\small\text{(f)}} &
\begin{minipage}[t]{0.21\textwidth}
  \includegraphics[width=\linewidth,height=0.195\textheight]{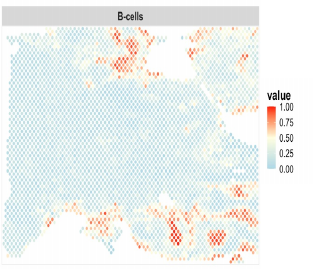}
\end{minipage}
\raisebox{0.12\textheight}{\small\text{(g)}} &
\begin{minipage}[t]{0.21\textwidth}
  \includegraphics[width=\linewidth,height=0.195\textheight]{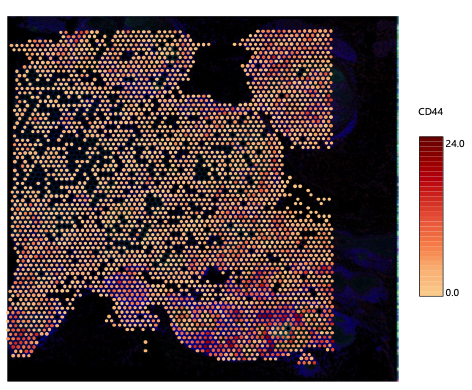}
\end{minipage}
\raisebox{0.12\textheight}{\small\text{(h)}} &
\begin{minipage}[t]{0.21\textwidth}
  \includegraphics[width=\linewidth,height=0.195\textheight]{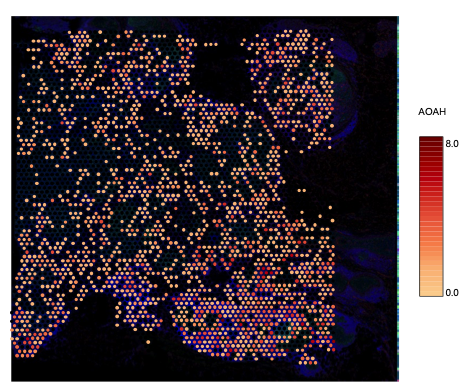}
\end{minipage}
\end{tabular}
		\caption{(a) Breast cancer tissue image. (b) Spatial distribution of the log-transformed CD8A protein expression. 
        (c) The posterior partition sample selected by the R package \texttt{salso}. 
        (d) Spatial distribution of the log-transformed summed expression of cancer marker proteins (normalized). Red colors indicate higher expression levels of tumor markers and thus a larger number of tumor cells. (e) Spatial distribution of T cell proportion. (f) Spatial distribution of B cell proportion. (g) Spatial distribution of CD44 gene expression. (h) Spatial distribution of AOAH gene expression.
        }
		\label{FIG:STcluster}
\end{figure}

The distribution of CD8A protein expression exhibits a long right tail; therefore, we apply a log transformation for normalization. Figure \ref{FIG:STcluster}(b) shows the spatial distribution of the transformed CD8A protein expression, with notably lower expression observed in the middle left region compared to other areas. Subsequently, both genes and CD8A protein expression are standardized to have a mean $0$ and standard deviation $1$. To reduce dimensionality, we first perform sure independence screening \citep{SIS} on all genes, followed by Lasso regression \citep{tibshirani1996regression}, with hyperparameters selected by cross-validation. This pre-screening procedure results in a set of 234 genes potentially associated with CD8A protein expression. 

Setting the screened $234$ genes as $\tmmathbf{x}_{\dagger}(\cdot)$ and the log-transformed CD8A protein as $y(\cdot)$, we apply our model under the VS setting. We apply the same method as in  Section \ref{SEC:simu} to choose $\{K,\alpha (\cdot),\lambda\}$, with constants $(c_b,c_a,c_p)$ selected from the grid $(2,2.5,3,3.5,4)\times(0.005,0.0075,0.01,0.025,0.05)\times(0.001,0.0025,0.005,0.0075,0.01)$ based on WAIC criteria. We set $q_{\tmop{max}} = 50$, $k_{\tmop{max}} = 15$, and $\sigma$ equal to the standard deviation of $\{y(\tmmathbf{s}_i)\}_{i=1}^{n}$. We run $20000$ MCMC iterations with a burn-in period $15000$ and a thinning parameter $5$.




Figure \ref{FIG:STcluster}(c) shows the posterior partition sample selected by the R package \texttt{salso} \citep{salso}.
Our model identifies three spatial clusters: Cluster 2 occupies the upper right (green), Cluster 3 occupies the bottom and most lower right (blue), while Cluster 1 covers the rest, the most left part (red).
We observe that Clusters 2 and 3 correspond to the darker region in Figure~\ref{FIG:STcluster}(a), while Cluster 1 corresponds to the lighter region.

To evaluate our partitioning results, we incorporate two sources of external information. First, we compute the summed expression levels of EPCAM, KRT5, PCNA, BCL2, VIM, and SDC1, proteins commonly used as breast cancer markers, all of which are available in the same SMO dataset. The spatial distribution of the summed expression is shown in Figure~\ref{FIG:STcluster}(d), where we observe that Cluster 1 is predominantly tumor-rich, while Clusters 2 and 3 contain scattered tumor islands. Second, we use an external scRNA-seq reference \citep{Wu2021} and deconvolute immune cell (T cell and B cell) type proportions using CARD~\citep{Ma2022}, and the results are shown in Figures~\ref{FIG:STcluster}(e-f). We observe that immune cells are mainly localized in Clusters 2 and 3, highlighting the presence of tertiary lymphoid structures, hubs of T and B cells that are highly related to anti-tumor immune response~\citep{Helmink2020},
in these areas.



Table~\ref{Tab:geneselection} reports the genes with posterior inclusion probabilities (IP) greater than 0.1, along with the corresponding posterior mean regression coefficients for each cluster.
These selected genes are cluster-specific and represent key components of the localized microenvironment of CD8$^{+}$ T cells, highlighting distinct CD8$^{+}$ T cell niches within each cluster region.
Most of the selected genes are positively correlated with the presence of CD8$^{+}$ T cells, while a few show negative correlations. The sets of selected genes for Cluster 2 and Cluster 3 largely overlap but exhibit slight differences, suggesting a potential functional shift in CD8$^{+}$ T cell states between the two regions.

\begin{table}[htbp]
\centering
\caption{Genes selection results}
\label{Tab:geneselection}
\resizebox{\textwidth}{!}{
\begin{tabular}{cccccccccc}
\hline
                      & \multicolumn{8}{c}{Genes} \\
                      \hline
Cluster 1             & TMSB4X   & CCL5  &  CYB5A     &  IKZF1&IFI27&IL7R&&& \\
IP (coefficients) & 1.00 (0.88)      & 0.96 (0.18)    &     0.69 (-0.13)  &0.65 (0.23) &0.25 (-0.05)& 0.21 (0.05) &&&\\
Cluster 2             & IL7R   & CCL5   & ZAP70  & SCD&KLRK1&GBP5&NLRC5&WASHC5&  \\
IP (coefficients) & 1.00 (0.34)      & 0.98 (0.21)    & 0.87 (0.18)     &0.60 (-0.16)&0.52 (0.08)&0.45 (0.06)&0.18 (0.03) &0.16 (-0.06)& \\
Cluster 3             & CCL5   & KLRK1   & SCD  & NLRC5&CD44&IL7R&AOAH&&  \\
IP (coefficients) & 1.00 (0.20)      & 0.94 (0.12)    & 0.52 (-0.11)     &0.41 (0.05)&0.34 (0.04)&0.33 (0.04)&0.10 (0.01)&&\\
\hline
\end{tabular}
}
\end{table}

We further examine the results in Table~\ref{Tab:geneselection}. Notably, all three clusters select CCL5 
and IL7R. CCL5 is a chemokine that recruits immune cells, including T cells, NK cells, and dendritic cells, to sites of immune activity, serving as a hallmark of the CD8$^{+}$ T cell niche. IL7R is a common marker for both CD4$^{+}$ and CD8$^{+}$ memory T cells. The inclusion of both genes is consistent with existing knowledge of genes correlated with CD8A expression.

Cluster 1 represents the CD8$^{+}$ T cell niche in the tumor region, especially at the tumor boundary, resembling a ``war zone" of T cell immune activity. 
Immune-associated genes such as TMSB4X, CCL5, IKZF1, and IL7R show positive correlations, suggesting the presence of early-activated or memory-like CD8$^{+}$ T cells. 
TMSB4X is involved in cytoskeletal remodeling and is expressed in activated T cells, while IKZF1 plays a key role in T cell differentiation.
In contrast, CYB5A and IFI27 exhibit a negative correlation with the CD8A protein expression, and both genes are known markers overexpressed in breast cancer: CYB5A, 
is associated with ER+ breast cancer and may reflect metabolically active tumor cells, whereas IFI27 is frequently upregulated in proliferative tumors. 
The coexistence of immune- and tumor-associated signals in this region suggests an active battleground between immune response and tumor evasion, where CD8$^{+}$ T cells may be mounting an attack while facing suppressive pressures from neighboring tumor cells.

Cluster 2 exhibits a gene expression profile consistent with a cytotoxic CD8$^{+}$ T cell niche in the stroma region with lymphoid structures. 
Positively associated genes (IL7R, CCL5, ZAP70, KLRK1, GBP5, NLRC5) indicate T Cell Receptor (TCR) signaling mediated T cell activation:
In addition to IL7R and CCL5, 
ZAP70 and NLRC5 are essential for TCR-mediated antigen recognition; KLRK1 (NKG2D) is a cytotoxic receptor and GBP5 marks interferon response.
Together, these selected genes suggest a microenvironment where CD8$^{+}$ T cells are cytotoxic and actively responding to tumor antigens, consistent with the function of tertiary lymphoid structures.
Unlike Cluster 1, 
Cluster 2 shows negative correlations with two different tumor-associated genes, SCD and WASHC5, reflecting distinct tumor transcriptomic states in this region.


The selected genes in Cluster 3 are largely similar to those in Cluster 2 with some subtle differences, possibly due to the greater abundance of tumor islands in Cluster 3 compared to Cluster 2, as shown in Figure~\ref{FIG:STcluster}(d). Compared to Cluster 2, Cluster 3 further includes positively correlated genes CD44 and AOAH, markers of memory-like T cells and myeloid antigen-presenting cells, respectively, suggesting an active myeloid-mediated antigen presentation in this region. 
This process enables myeloid cells to present tumor antigens to CD8$^{+}$ T cells, 
facilitating tumor recognition and subsequent cytotoxic response.
Figures~\ref{FIG:STcluster}(g-h) show the spatial distribution of CD44 and AOAH gene expression levels, respectively, with higher expression observed in the lower right region, which further supports the hypothesis of active myeloid-mediated antigen presentation in Cluster 3. 

\section*{Conclusion}\label{sec-conc}
We propose an LSFS framework for large-scale spatial data that simultaneously enables domain partition and feature selection. The domain partition prior uses a blocking strategy that reduces the infinite domain partition space to the finite blocking partition space, thereby ensuring feasibility and scalability. The notion of “feature” includes both
covariates and functional bases, enabling both local variable selection and local basis selection. We derive hyperparameter conditions, under which the consistency
theories of both the domain partition and feature selection are
established in the VS setting. To address the computational challenges of jointly sampling domain partitions and selected features, we design an efficient informed RJ-MCMC algorithm. The application of our model to the SMO dataset reveals the tumor and lymphoid structures within breast cancer tissue, along with distinct T cell functionalities inferred from the selected genes.

Moving forward, there are several avenues for future research. First, the proposed model currently handles only a single response (i.e., one protein in the SMO dataset), whereas the raw dataset measures 35 proteins in total. Extending the framework to jointly model multiple proteins and their relationships with gene expression would provide valuable insights into protein–protein interaction structures. Second, the choice of hyperparameter strategies in both the SVS and NR settings requires further theoretical investigation. In the SVS setting, the nonparametric component introduces additional approximation bias, making it more challenging to establish consistent hyperparameter conditions. In the NR setting, although simulation results demonstrate improved performance compared with existing methods, a clear theoretical guideline for hyperparameter selection is still lacking. We leave these issues as important directions for future research.



\section*{Data Availability Statement}\label{data-availability-statement}

The SMO dataset \citet{10xgenomics_breastcancer} analyzed in Section \ref{SEC:realdata} is available at \url{https://www.10xgenomics.com/datasets/gene-and-protein-expression-library-of-human-breast-cancer-cytassist-ffpe-2-standard}



\newpage


\vspace{-5mm}  
\baselineskip=14pt 
\singlespacing 
\begingroup
    \setstretch{0.5}
    \bibliographystyle{apalike}
    {\small
    \bibliography{ref}}
\endgroup



\end{document}